# Probing frontier orbital energies of {Co₉(P₂W₁₅)₃} polyoxometalate clusters at molecule–metal and molecule–water interfaces

Xiaofeng Yi,†,‡ Natalya V. Izarova,† Maria Stuckart,†,‡ David Guérin,§ Louis Thomas,§ Stéphane Lenfant,§ Dominique Vuillaume,§,* Jan van Leusen,‡ Tomáš Duchoň,‖ Slavomír Nemšák,†,¶ Svenja D. M. Bourone,‡ Sebastian Schmitz‡ and Paul Kögerler†,‡,*

† Jülich-Aachen Research Alliance (JARA-FIT) and Peter Grünberg Institute 6, Forschungszentrum Jülich, D-52425 Jülich, Germany

‡ Institute of Inorganic Chemistry, RWTH Aachen University, D-52074 Aachen, Germany

§ Institute of Electronics, Microelectronics and Nanotechnology, CNRS, University of Lille, 59652 Villeneuve d'Ascq, France

‖ Faculty of Mathematics and Physics, Charles University, 18000 Prague, Czech Republic

¶ BESSY-II, Helmholtz-Zentrum Berlin, D-12489 Berlin, Germany

**ABSTRACT:** Functionalization of polyoxotungstates with organoarsonate co-ligands enabling surface decoration was explored for the triangular cluster architectures of the composition $[Co^{II}_9(H_2O)_6(OH)_3(p\text{-}RC_6H_4As^VO_3)_2(\alpha\text{-}P^V_2W^{VI}_{15}O_{56})_3]^{25-}$ ({Co₉(P₂W₁₅)₃}, R = H or NH₂), isolated as Na₂₅[Co₉(OH)₃(H₂O)₆(C₆H₅AsO₃)₂(P₂W₁₅O₅₆)₃]·86H₂O (**Na-1**; triclinic, $P\bar{1}$, $a = 25.8088(3)$ Å, $b = 25.8336(3)$ Å, $c = 27.1598(3)$ Å, $\alpha = 78.1282(11)°$, $\beta = 61.7276(14)°$, $\gamma = 60.6220(14)°$, $V = 13888.9(3)$ Å, $Z = 2$) and Na₂₅[Co₉(OH)₃(H₂O)₆(H₂NC₆H₄AsO₃)₂(P₂W₁₅O₅₆)₃]·86H₂O (**Na-2**; triclinic, $P\bar{1}$, $a = 14.2262(2)$ Å, $b = 24.8597(4)$ Å, $c = 37.9388(4)$ Å, $\alpha = 81.9672(10)°$, $\beta = 87.8161(10)°$, $\gamma = 76.5409(12)°$, $V = 12920.6(3)$ Å, $Z = 2$). The axially oriented *para*-aminophenyl groups in **2** facilitate the formation of self-assembled monolayers on gold surfaces, and thus provide a viable molecular platform for charge transport studies of magnetically functionalized polyoxometalates. The title systems were isolated and characterized in the solid state and in aqueous solutions, and on metal surfaces. Using conducting tip atomic force microscopy (C-AFM), the energies of {Co₉(P₂W₁₅)₃} frontier molecular orbitals in the surface-bound state were found to directly correlate with cyclic voltammetry data in aqueous solution.

## INTRODUCTION

Highly energy-efficient molecular spintronics is emerging as a rapidly growing field due to the prospects of the combined exploitation of molecular charge and spin states.[1] Polyoxometalates (POMs) as discrete nanoscale metal-oxo clusters able to incorporate magnetic centers in their structures offer a number of advantages for creating molecular spintronic devices, such as thermal and redox stability coupled with large structural diversity and tunability of the magnetic properties.[2] One of the key challenges to fabricate molecular devices is control over molecular anchoring on the surfaces of metallic electrodes in terms of spatial molecular orientation as well as the degree of electronic interactions between the metallic surface states and the molecular orbitals.[3] As such, understanding the electronic consequences caused by the interactions of magnetically functionalized POMs with metal substrates is essential to their eventual use in molecular spintronics. We here elaborate a POM-based model system that allows us to assess and compare the molecular frontier orbitals that are accessible in molecular charge-transport measurements for a surface-adsorbed POM via independent methods, namely transient voltage spectroscopy and cyclic voltammetry.

A possible step towards this goal relies on the pre-functionalization of a magnetic molecule with "glue groups" (*e. g.* –SH, –N₂⁺, –NH₂ *etc.*) that covalently bind, or chemisorb, to a specific surface. In recent years it was demonstrated that such groups can be introduced into POM species by attachment of various organic ligands, *e. g.* alkoxides, siloxanes, organo(bis)phosphonates and -arsonates,[4-6] although examples of functionalized magnetic POMs are still scarce,[5] despite the prospect that POM functionalization with glue groups has already enabled well-ordered patterning of various surfaces.[6] Here we explore a novel surface anchoring mode (organoamino group–Au surface) in an approach to render magnetically functionalized POMs accessible to charge transport experiments in distinct environments, in solution as well as in surface-adsorbed monolayers. For the design of our target molecules we exploited the fact that organoarsonates on one hand can provide a robust tetrahedral arsonate site that often can be readily integrated as a part of a magnetic core of transition metal ions, *e. g.* to replace phosphate groups possessing a terminal oxygen (see [7] for some examples of POMs containing such HPO₄⁻ groups). On the other hand, starting with phenylarsonate, as an easily accessible model ligand, one can further introduce various functional substituents to the phenyl ring (e. g. –NH₂ groups in *meta*- or *para*-positions). Until now no magnetic organoarsonate-polyoxotungstate derivatives have been reported, although several magnetic bisphosphonates-containing POMs are known,[5] e. g. [{(B-α-PW₉O₃₄)Co₃(OH)(H₂O)₂(Ale)₂}₂Co]¹⁴⁻ (H₅Ale = (⁺H₃N(CH₂)₃)(OH)C(PO₃H₂)₂) that was shown to exhibit single-molecule magnet features and is composed of two {B-α-PW₉O₃₄Co₃} subunits connected via an additional Co^II center as well as two alendronate ligands.[5a]

Our efforts resulted in isolation of two novel species with general formula [Co^II₉(H₂O)₆(OH)₃(*p*-RC₆H₄AsO₃)₂(α-P₂W₁₅O₅₆)₃]²⁵⁻, where R is either H (**1**) or NH₂ (**2**), which were crystallized as hydrated sodium salts Na₂₅[Co₉(H₂O)₆(OH)₃(*p*-RC₆H₄AsO₃)₂(α-P₂W₁₅O₅₆)₃]·86H₂O (**Na-1** and **Na-2**, respectively) and characterized in the solid state and aqueous solutions. We also carried out surface deposition studies on bare gold surfaces for the amino-terminated polyanion **2**, which are of particular interest as to date the amine-containing molecule self-assembled monolayers (SAMs) on Au were less studied in comparison with thiol-containing SAMs.[8]

## RESULTS AND DISCUSSION

**Synthesis.** The polyanions were prepared by reacting of CoCl₂, phenylarsonic (for **1**) or *p*-arsanilic (for **2**) acid and the tri-lacunary Wells-Dawson type polyoxotungstate salt Na₁₂[α-P₂W₁₅O₅₆]·24H₂O[9] in 0.66 M CH₃COOH / CH₃COONa buffer solution (pH 5.2) at 60 °C for four days. Plate-like brown crystals of Na.₁₆–ₓHₓ[(H₂O)₂Co₄(α-P₂W₁₅O₅₆)₂]·nH₂O (**Na-3**), based on the well-known sandwich-type polyanion [(H₂O)₂Co₄(α-P₂W₁₅O₅₆)₂]¹⁶⁻ (**3**),[10] form as the first product during evaporation of the reaction solution and should be repeatedly removed by filtration. Further evaporation of the obtained filtrate within several days leads to pink needle-like (or elongated plate-like) crystals of **Na-1** or **Na-2**. The crystalline materials of **Na-1** and **Na-2** should be collected within 1 to 2 days after formation to prevent their contamination with **Na-3** side-product. The use of any larger amounts of Co^II ions (than specified in the synthetic procedures) during the synthesis of **Na-1** or **Na-2** leads to the presence of Co^II as countercations and thus should be prevented. The influence of other synthetic reaction parameters on the formation of **1** and **2** is discussed after the structural description.

**Crystal structures.** Single-crystal X-ray structural analysis of **Na-1** and **Na-2** revealed that polyanions **1** and **2** display a similar molecular structure based on a {Co^II₉(H₂O)₆(OH)₃(*p*-RC₆H₄AsO₃)₂} core (Co₉L₂, where L = C₆H₅AsO₃²⁻ for **1** and *p*-H₂NC₆H₄AsO₃²⁻ for **2**, Fig. 1c), stabilized by three [α-P₂W₁₅O₅₆]¹²⁻ POT moieties (Fig. 1a, b).

Alternatively, the core structure of **1** and **2** can be viewed as a C₃ₕ-symmetric trimer of {Co₃(H₂O)₂P₂W₁₅O₅₆} units (that can be considered as tri-substituted {M₈P₂O₅₆} Wells-Dawson-type phosphotungstates) linked together by three OH groups and two phenylarsonate or *p*-arsanilate ligands. **1** and **2** represent equilateral triangular structures with side lengths of ca. 2.2 nm and a thickness (maximum extension of the organoarsonates groups) of ca. 1.7 nm.

Each Co^II ion in the {Co₃(H₂O)₂P₂W₁₅O₅₆} building block resides in an octahedral coordination environment and coordinates to one μ₄-O (3Co, P) of the inner phosphate group (Co–O: 2.166(14) – 2.268(13) Å for **1**, 2.156(14) – 2.239(14) Å for **2**) and two μ₂-O (Co, W) atoms (Co–O: 2.014(14) – 2.109(12) Å for **1**, 2.014(14) – 2.100(14) Å for **2**) of a {P₂W₁₅} subunit. One of the three Co^II ions additionally coordinates two μ₃-O, linking it with two other Co^II centers and the As^V center of the phenylarsonate or *p*-arsanilate ligand (Co–O: 2.038(13) – 2.119(13) Å for **1**, 2.048(14) – 2.093(14) Å for **2**) as well as a μ₃-OH group that connects it to two Co^II ions of the neighboring {Co₃(H₂O)₂P₂W₁₅O₅₆} unit (Co–O: 2.041(13) – 2.146(13) Å for **1**, 2.048(14) – 2.093(14) Å for **2**).

The other two Co^II ions bind to a terminal aqua ligand (Co–O: 2.111(15) – 2.135(14) Å for **1**, 2.087(16) – 2.157(15) Å for **2**), one above-mentioned μ₃-oxygen of the phenylarsonate/*p*-arsanilate and a μ₃-OH group linking them to each other and to a Co^II of the third {Co₃(H₂O)₂P₂W₁₅O₅₆} moiety (Fig. 1a). The protonation sites (bridging OH and terminal H₂O ligands) are confirmed by bond valence sum calculations (Tables S1, S2). The P–O and W–O bonds in **1** and **2** are typical for POTs. As–O bonds amount to 1.659(12) to 1.685(12) Å for **1** and 1.680(13) to 1.704(15) Å for **2**, while As–C bonds are 1.89(2) – 1.98(13) Å (**1**) and 1.93(2) Å (**2**). The two phenylarsonate/*p*-arsanilate ligands in **1**/**2** are located on the opposite sides of the {Co₉(H₂O)₆(OH)₃P₂W₁₅O₅₆)₃} assembly, with their (*p*-amino)phenyl groups oriented along the C₃ axis of the inorganic core structure (Fig. 1b,c). These organic groups also eliminate all symmetry elements of polyanions **1** and **2**; the dihedral angle between the two phenyl rings is 81.8(7)° in **1** and 39.4(7)° in **2**. Overall, the structure of the {Co₉(H₂O)₆(OH)₃Co₄(α-P₂W₁₅O₅₆)₃} assemblies in **1** and **2** is reminiscent of [Co₉(H₂O)₆(OH)₃(HPO₄)₂(α-P₂W₁₅O₅₆)₃]²⁵⁻ (**4**) reported several years ago by the Cronin group,[7e] where two phosphate groups are present instead of phenylarsonate or *p*-arsanilate. The central {Co₉(H₂O)₆(OH)₃(HPO₄)₂} core is also present in [Co₉(H₂O)₆(OH)₃(HPO₄)₂(PW₉O₃₄)₃]¹⁶⁻ polyanions[7a-c] that were shown to catalyze heterogeneous water oxidation[11a-c] while stabilized by Keggin-type {PW₉O₃₄}-type POTs, along with several other Co(II)-based POTs.[11d]

We note that already back in 1984, Weakley predicted the possibility to replace terminal oxygen in each of the two external phosphate groups by alkyl or aryl groups for POM functionalization.[7a] Our findings support a general strategy that uses organoarsonate ligands for functionalization of magnetic polyanions incorporating tetrahedral phosphate or arsenate groups with terminal oxo/hydroxo groups. We also explored similar reactions with organophosphonates (*e. g.* phenylphosphonate), this however only resulted in the formation of the hydrated sodium salts of the known phosphate-based POMs **4**, Na$_x$H$_{25-x}$[Co$_9$(H$_2$O)$_6$(OH)$_3$(HPO$_4$)$_2$($\alpha$-P$_2$W$_{15}$O$_{56}$)$_3$]·$n$H$_2$O (**Na-4**).[12] Formation of **Na-4** was also observed by reacting CoCl$_2$, phenylarsonate (or *p*-arsanilate) and {P$_2$W$_{15}$} in 1M LiCl at higher pH values (8–10) or in Na$_2$CO$_3$ / NaHCO$_3$ buffer solution (pH 9.4). We hypothesize that high pH of the reaction medium leads to partial {P$_2$W$_{15}$} decomposition and the release of free phosphate required for formation of **4**. Also addition of dimethylammonium chloride (DMACl) or CsCl to our reaction mixtures for preparation of **1** and **2** results in crystallization of polyanions **4** rather than DMA or Cs salts of the desired products. Thus, pH and counterion size play a crucial role in the isolation of **1** and **2** over non-functionalized **4**.

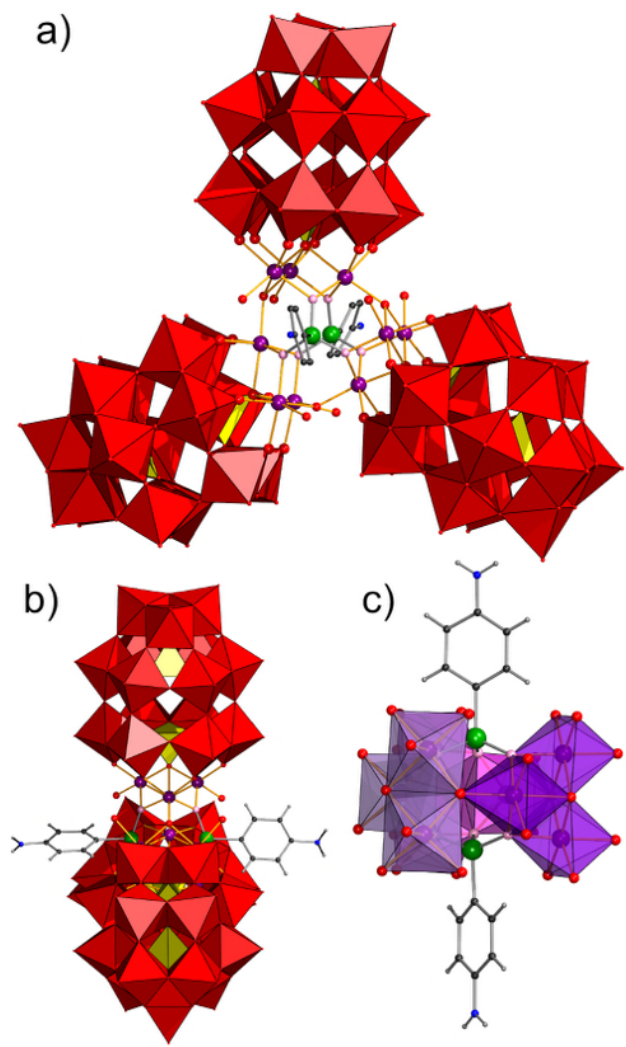

a) b) c)

**Figure 1.** Structure of polyanion **2** in front (a; H atoms omitted for clarity) and side (b) views. (c) Structure of the {Co$_9$(*p*-ars)$_2$} core. Color code: WO$_6$: red octahedra, PO$_4$: yellow tetrahedral; As: green, Co: purple, C: black, N: blue, H: gray, O: red spheres; O sites belonging to O$_3$As groups are shown in pink, bonds in the phenylarsonate or *p*-arsanilate groups are in gray, Co–O bonds in orange.

As expected, IR spectra of **Na-1** and **Na-2** (Fig. S1) are similar and exhibit a set of three peaks characteristic for P–O vibrations at 1084, 1042 and 1011 cm$^{-1}$ for **Na-1** and 1086, 1042 and 1009 cm$^{-1}$ for **Na-2**, comparable to the peaks in the IR spectrum of Na$_{12}$[$\alpha$-P$_2$W$_{15}$O$_{56}$]·24H$_2$O at 1130, 1086 and 1009 cm$^{-1}$. The disappearance of the band at 1130 cm$^{-1}$ and appearance the band at 1042 cm$^{-1}$ are in agreement with coordination of the O atom of PO$_4$ group at the lacunary site of {P$_2$W$_{15}$} ligand by three Co$^{II}$ ions in **1** and **2**. The bands at 933 (**Na-1**) and 932 cm$^{-1}$ (**Na-2**) correspond to W=O vibrations. Peaks at 880, 806, 731, 598, 525 and 459 cm$^{-1}$ (**Na-1**) and 881, 806, 725, 600, 521 and 457 cm$^{-1}$ (**Na-2**) are associated with W–O–W, W–O–Co and W–O–P bond vibrations. The remarkable shift of these bands compared to W–O–W bands in non-coordinated {P$_2$W$_{15}$} (see Fig. S1) is consistent with the coordination of Co$^{II}$ to {P$_2$W$_{15}$} in **Na-1** and **Na-2**. The characteristic As–O bands (800 – 815 cm$^{-1}$)[13] overlap with W–O–W/Co/P modes (strong band at 806 cm$^{-1}$). Overlap also affects As–C, C–C and C–H vibrations of the organic moieties in **Na-1** and **Na-2** in the POT region. However, C–C and C–H vibrations result in weak bands between 1600 and 1100 cm$^{-1}$ (Fig. S1). Characteristic to **Na-1** is the sharp C–N band at 1352 cm$^{-1}$.

We have additionally performed ATR FT-IR measurements on saturated solution of **Na-2** in H$_2$O in comparison to solid **Na-2** sample in KBr. The good match of the spectra (Fig. S3) suggests solution stability of polyanions **2** in aqueous medium, at least within the duration of the measurement.

**Thermogravimetrical analysis**. The thermal stability of **Na-1** and **Na-2** was investigated in the range of 25 – 900 °C under N$_2$ atmosphere. The TGA curves of both compounds are similar (Figs. S4, S5) and exhibit a major mass loss in several consecutive non-resolved steps up to 300 °C due to the release of 86 lattice water molecules per formula unit (10.6 % observed *vs.* 10.8 % calcd. for **Na-1**, 10.3 % obs. *vs.* 10.8 % calcd. for **Na-2**). Several additional steps in the 300 – 690 °C range are attributed to the loss of six coordinated water molecules and three hydroxo ligands, combined with decomposition and removal of phenyl (**Na-1**) or *p*-aminophenyl (**Na-2**) groups of the *p*-RC$_6$H$_4$AsO$_3^{2-}$ ligands as well as with O$_2$ release due to reduction of As$^V$ ions (2.5 % obs. *vs.* 1.7 % calc. for **Na-1** and 2.6 % obs. *vs.* 1.8 % calc. for **Na-2**). Additional decrease in mass between 700 – 800 °C may stem from loss of volatile arsenic oxide (*e.g.* as incomplete release of 0.5 As$_4$O$_6$ per formula unit: 0.8 % exp. *vs.* 1.4 % calc. (**Na-1**) and 0.6 % obs. *vs.* 1.4 % calc. (**Na-2**)). The total mass loss at 900 °C is 14.0 % for **Na-1** and 13.8 % for **Na-2**.

**Magnetochemical analysis**. The magnetic data of **Na-1** and **Na-2** are shown as $\chi_m T$ *vs.* $T$ and $M_m$ *vs.* $B$ curves in Fig. 2 (molar magnetic susceptibility $\chi_m$, molar magnetization $M_m$, temperature $T$, and magnetic field $B$). At 290 K, the $\chi_m T$ value of **Na-1** is 25.98 cm$^3$ K mol$^{-1}$ at 0.1 T, i.e. in the range of 20.81–30.43 cm$^3$ K mol$^{-1}$ expected for nine non-interacting, octahedrally coordinated high-spin Co$^{II}$ centers.[14] Upon lowering $T$, $\chi_m T$ decreases, initially slowly ($T \geq 180$ K) and subsequently rapidly, down to 3.37 cm$^3$ K mol$^{-1}$ at 2.0 K. At this temperature, the molar magnetiza-tion $M_m$ is almost linear with the applied field up to ca. 2 T. For higher fields, the magnetization subsequently increases with continuously decreasing slope yielding 8.1 $N_A$ $\mu_B$ at 5.0 T, well below the expected saturation value of 30.0–36.3 $N_A$ $\mu_B$ for nine non-interacting Co$^{II}$ centers. Both curves thus reveal predominant antiferromagnetic exchange interactions between the nine Co$^{II}$ centers in **1**. While the rapid decrease of $\chi_m T$ upon cooling below ≈ 100 K, is, for the most part, due to these exchange interactions, the $\chi_m T$ vs. T curve is also effected by the ligand field of each single Co$^{II}$ center causing a similar contribution that distinctly deviates from the spin-only behavior in the range 2.0–180 K. This is due to the thermal depopulation of the energy states originating from the $^4T_{1g}$ ground multiplet of the $^4F$ ground term that is further split by spin-orbit coupling contributions, in particular due to mixing with the states originating from the exited $^4T_{1g}(^4P)$ multiplet.[15]

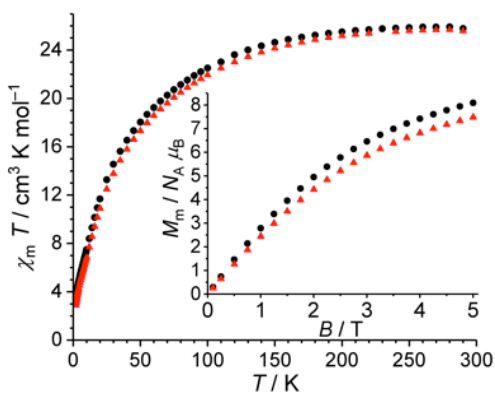

**Figure 2.** Temperature dependence of $\chi_m T$ at 0.1 Tesla of **Na-1** (back circles) and **Na-2** (red triangles). Insets: molar magnetization $M_m$ vs. applied field $B$ at 2.0 K.

The magnetic data of **Na-2** and of **Na-1** match almost perfectly. Taking into account the structure of both compounds, the slight differences of the characteristic values (**Na-2**: at 0.1 T, $\chi_m T$ = 25.75 cm$^3$ K mol$^{-1}$ at 290 K, and 2.91 cm$^3$ K mol$^{-1}$ at 2.0 K; $M_m$ = 7.6 $N_A$ $\mu_B$ at 5.0 T and 2.0 K) are potentially due to the slightly different exchange coupling mediated by the organoarsonate ligands, but the differences might also be caused by minor paramagnetic impurities. Nevertheless, the conclusions drawn from the magnetochemical analysis of **Na-1** are the same for **Na-2**.

A quick comparison of the magnetic data of **Na-1** and **Na-2** to **Na-4** [7e] reveals extensive resemblance. The shape of the $\chi_m T$ vs. T and $M_m$ vs. B curves are very similar, while their magnitudes are different. This may be caused by the different ligands (organoarsonate (**Na-1**, **Na-2**) vs. phosphate groups (**Na-4**)). The magnitude of deviation (about 25 %) is, however, surprisingly large which rather points to an uncertainty in the molar mass (as scaling factor), i.e. a different amount of a diamagnetic component such as crystal water within the measured sample. The samples used for magnetic measurements thus were subsequently characterized by TGA, however, the results are the same as presented above. In addition, compounds **Na-1**, **Na-2** and **Na-**

**4** do not show out-of-phase ac susceptibility signals down to 1.8 K and up to 1000 Hz.

**UV-Vis spectroscopy.** Aqueous solutions of **Na-1** and **Na-2** (Figs. S6, S8) show similar absorption spectra that exhibit a strong maximum at around 200 nm (215 nm for **Na-1** and 195 nm for **Na-2**) followed by an absorption maximum at around 260 nm (which is well-resolved for **Na-1** and is overlapped with the first absorption peak for **Na-2**) in the UV light area and a less intense absorption maximum at 532 nm in the visible light area. The spectrum remains unchanged for at least 18 h confirming the short-term stability of polyanions in aqueous solutions (Fig. S9) in line with the conclusions obtained from the Diamond ATR-FTIR measurements.

**Electrochemical studies.** We have recorded cyclic voltammograms for 0.7 mM **Na-1**, **Na-2**, **Na-3** and **Na-4** solutions in 0.5 M CH$_3$COONa buffer (pH 4.8). The electrochemical behavior of the three Co$_9$-based POMs (**1**, **2** and **3**) is very similar. The cyclic voltammograms for these species exhibit four redox couples attributed to reduction and re-oxidation of the W$^{VI}$ centers of the POT ligands between −0.50 and −1.05 V vs. Ag / AgCl (Figs. 3, S10, Table 1). At potentials below −1.05 V a reduction of solvent occurs, coupled with formation of a film on the glassy carbon electrode. In comparison to the electrochemical activity of the sandwich-type polyanions **3** in the same medium, exhibiting three well-defined redox waves before hydrogen evolution (Table 1, Fig. S11), reduction of W$^{VI}$ centers in the Co$_9$-based species takes place at more negative potentials showing higher redox stability of the latter. The peak currents for the redox processes in **1 − 4** are proportional to the square root of the scanning rate, which is characteristic for diffusion-controlled electrode reactions. At higher pH (6.4) all redox waves are shifted towards more negative potentials (Fig. S12) indicating that the reduction of tungsten(VI) ions is coupled with proton transfer, as it is common for POTs. Correspondingly, only three redox waves are accessible for **2** in the pH 6.4 medium before hydrogen evolution. No reversible redox waves associated with Co$^{II}$ oxidation in the {Co$_9$L$_2$} core could be observed in the positive potential range at both pH 4.8 and 6.4.

The CV curves of **1**, **2** and **4** remain unchanged for several hours, however after one day an additional shoulder centered at around −0.50 V appears, while there are also other slight changes in relative intensities and shape of the other redox waves. The shoulder at −0.50 V looks very similar to the first redox wave observed for **3** in the same medium (Fig. S13). These changes are accompanied by a color change of the POMs solutions from pink to light-brown. These observations imply partial decomposition of the {Co$_9$L$_2$(P$_2$W$_{15}$)$_3$} POMs in 0.5 M sodium acetate solutions (pH 4.8) following by formation of the {Co$_x$Na$_{4−x}$(P$_2$W$_{15}$)$_2$} species, this is also in agreement with the presence of **Na-3** as a common side product during the synthesis of **Na-1** and **Na-2**.

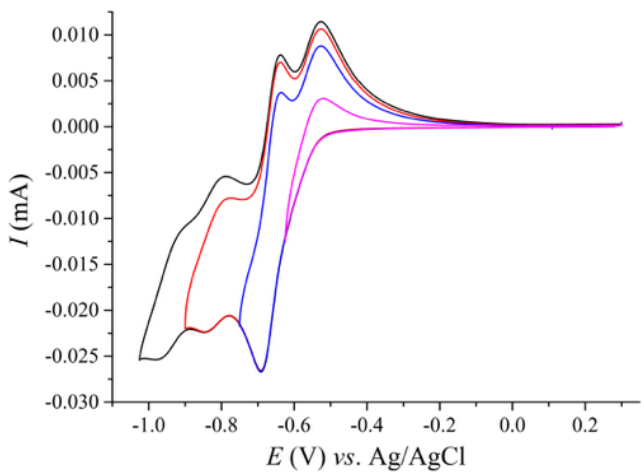

**Figure 3.** Room-temperature cyclic voltammograms of 0.7 mM solution of **Na-1** in 0.5 M CH$_3$COONa buffer (pH 4.8) with different negative potential limits (−0.625 V, −0.75 V, −0.95 V and −1.025 V) at scan rate 20 mV/s.

**Table 1. Electrochemical data for the polyanions 1 − 4 in 0.5 M CH$_3$COONa buffer (pH 4.8). The potentials are given *vs*. Ag/AgCl reference electrode at 20 mV/s scan rate.**

| POM | $E_{1/2}$, I, V | $E_{1/2}$, II, V | $E_{1/2}$, III, V | $E_{1/2}$, IV, V |
|-----|-------------|--------------|---------------|---------------|
| 1 | −0.57±0.04 | −0.67±0.03 | −0.82±0.03 | −0.94±0.03 |
| 2 | −0.56±0.07 | −0.67±0.04 | −0.82±0.03 | −0.94±0.03 |
| 4 | −0.58±0.07 | −0.67±0.05 | −0.80±0.02 | −0.92±0.03 |
| 3 | −0.50±0.08 | −0.70±0.06 | −0.89±0.06 | − |

## Formation and characterization of self-assembled monolayer (SAM).

We investigated the electron transport properties of a molecular device based on the amine-terminated polyanions **2**. For this purpose, we prepared SAMs of **2** on flat bare template-stripped gold surface (Au$^{TS}$) [17] and characterized it using ellipsometry, FT-IRRAS, X-ray photoemission spectroscopy (XPS), cyclic voltammetry (CV) and C-AFM measurements.

It was found that the incubation of the freshly prepared gold surface in the aqueous solution of **Na-2** (10$^{−3}$ M in DI water) during 24 hours leads to the formation of a layer with the thickness between 1.4 ± 0.2 and 2.0 ± 0.2 nm (measured on two batches, see experimental section), which is consistent with the expected thickness of the polyoxoanion monolayer in the most probable orientation on the gold surface shown on Fig. S14 (~1.7 nm according to the XRD data). The thickness of the obtained layer did not change after ultrasonic cleaning of the sample in water, indicating a high stability of the POM SAM.

The results of topographic AFM in contact mode (Fig. S15) further suggest the homogeneity of the obtained layer. The RMS roughness of 0.31 nm (root mean square roughness is defined here as the square root of the arithmetic mean of the squares of each z values measured on an AFM image) obtained by AFM is in a good agreement with a standard deviation of 0.2 nm observed by ellipsometry. This value is also well comparable with the RMS roughness of the template-stripped (TS) Au surface before grafting molecules measured at the same conditions, which is around 0.5 nm. However, we note that AFM images reveal some pinholes with a diameter of few tens of nm and apparent depth of about 1 nm (about half the total SAM thickness, albeit the exact value cannot be determined due to AFM tip convolution effect).

The obtained POM-covered surface was also characterized by CV. The three-electrode cell setup was used with a POM-covered gold surface as the working electrode, Pt wire as a counter electrode, and Ag/AgCl (saturated KCl) as a reference electrode (Fig. S16). The characteristic wave of the successive one-electron reduction process attributed to **2** on the gold surface (black line, Fig. S16) was identified at ca. −0.38 V *vs*. Ag/AgCl (saturated KCl), but it is not well defined and appears irreversible. We also perform CV measurements of the **Na-2** in the same electrolyte solution we used for CV studies of the POM-containing surface for comparison (red line, Fig. S16). For this measurements Pt wires acted as working and counter electrodes, whereas Ag/AgCl (saturated KCl) was used as a reference electrode.

The XPS experiments of the Na-2 SAM on gold (Fig. S17) show distinctive photoemission lines of W4f7/2 (36.2 eV) and W4f5/2 (38.4), As3d (42.0 and 45.5 eV), C1s (284.8 eV) and Co2p (~782 eV, almost invisible). The results are confirmed by XPS measurements of the Na-2 powder sample, which also reveal photoemission peaks of the other elements of interest (Fig. S18). Due to its low content and relatively low photoionization cross-sections, a very weak photoemission peak of N 1s (~401 eV) is present just at the detection limit of the experiment and hints at a broadening and splitting indicative of attachment of the amino group to the Au substrate. Doublets of P 2p (~136 eV) and Co 2p (~781 eV) for the powder sample are significantly stronger comparing to the Na-2 on Au, which is aligned with our expectations. A semi-quantitative analysis of the photoemission lines belonging to Co, Na, P, W and C was performed using the SESSA simulation package.[18] The analysis confirms the powder retains its stoichiometry. Photoelectron cross-sections values were taken from ref. 19, inelastic-mean free path was calculated using the TPP-2 formula.[20] Interestingly, the Na 1s peak was not observed for Na-2 SAM on gold, in contrast to the XPS data for bulk material (~1073 eV for Na-2 powder).

Additionally, a thin layer of **Na-2** immobilized on an Au surface from an aqueous solution was characterized by FT-IRRAS (Fig. S19). The measurement reveals several characteristic peaks that further supports uniform **Na-2** immobilization on the Au surface. Thus, the bands at 821 cm$^{−1}$ as well as at 919 cm$^{−1}$ are most likely vibrations of W–O–W, Co–O–W, P–O–W bonds, while that at 1108 cm$^{−1}$ corresponds to vibrations of the P–O bonds. They are slightly shifted as compared to the peaks detected by FT-IR mea-surements in transmission of **Na-2** powder (806, 881 cm⁻¹; and 1086 cm⁻¹, respectively). This shift is expected due to the different sample forms and measurement methods that lead to variant vibration frequencies.[21] The numerous rotational-vibrational transitions observed between 2000 cm⁻¹ and 1200 cm⁻¹ should be owing to the hydrophilic nature of **Na-2**, the high amount of crystal water present in this compound as well as the preparation of the sample from ultrapure water.

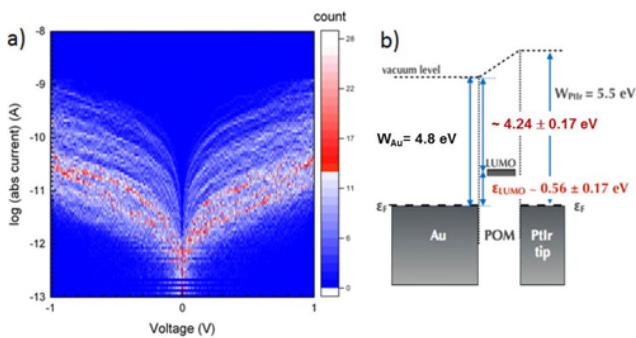

**Figure 4.** (a) 2D current histogram of 325 *I-V* curves measured by C-AFM (at a loading force of 30 nN) on the SAM of **2** chemically grafted on ultra-flat template stripped gold electrode (Au^TS). Voltages applied on the Au^TS electrode (C-AFM grounded). (b) Schematic energy diagram of the Au^TS / POM / PtIr C-AFM tip as deduced from both the *I-V* and CV measurements.

Current-voltage (*I-V*) curves were measured by electrically contacting the SAM by C-AFM. Figure 4a shows the 2D histograms of around three hundred of *I-V* curves (see experimental part). The *I-V* curves are dispersed over about three decades as shown by current histograms at a given bias (*e. g.* +0.9V and −0.9 V, Fig. S20). This dispersion is large when compared to similar C-AFM measurements of alkylthiol SAMs on Au (here about 3 decades *vs.* about one [22]) or π-conjugated molecules.[23] The *I-V* curves are symmetric with respect of the voltage polarity. The current histograms are fitted by two Gaussian peaks (Fig. S20), the mean values of the current histograms (Fig. S20) are *ca.* 1.82×10⁻¹¹ A / 1.6×10⁻¹⁰ A (two peaks) at 0.9 V and *ca.* 1.9×10⁻¹¹ A / 2.1×10⁻¹⁰ A at −0.9 V. The results have been reproduced in a second batch of SAMs (see Figs. S22-S24).

We can infer several possible origins for this large dispersion:

(1) Disorder in the SAMs. For example, in some cases multi-peaks and consequently a larger dispersion can be observed for disordered SAMs over two decades of current.[24] This large dispersion can also be due to the presence of "defects" (such as pinholes, see above and Fig. S15) with reduced thickness and thus larger current through the SAMs. Thus, the two peaks in the current histograms can be ascribed to zones of different molecular organizations in the SAMs.

(2) Intermolecular interaction (π-π interactions) between neighboring molecules which can broaden the conductance histograms.[25]

We also analyze the *I-V* curves with the TVS (transient voltage spectroscopy) method[26] to estimate the energy position (with respect to the Fermi energy of the electrodes) of the molecular orbital (HOMO or LUMO) involved in the charge-transport process (Fig. 4b). From the minima of the TVS plots (Fig. S21), we deduced (see Experimental Part) a LUMO at *ca.* (0.56±0.17) eV above the electrode Fermi energy (energy diagram Fig. 4b), or at (4.24±0.17) eV below the vacuum level (considering a work function of Au^TS at 4.8 eV). This later value is consistent with the LUMO position determined from CV on the SAM, the reduction peak at −0.38 V giving a LUMO at −4.2 eV below vacuum level (see equation in the Experimental Part). This consistency between *I-V* and CV measurements indicates a weak or moderate electronic coupling between the molecules and the Au surface (through the phenyl amine groups) on one side and through the C-AFM tip mechanical contact on the other side.

EXPERIMENTAL PART

**General methods and materials**. Reagents were used as purchased without further purification. Na₁₂[α-P₂W₁₅O₅₆]·24H₂O was obtained according to the reported procedure [9] starting from K₆[α-P₂W₁₈O₆₂]·14H₂O.[27] Elemental analysis results (ICP-OES and C, H, N) were obtained from Central Institute for Engineering, Electronics and Analytics (ZEA-3), Forschungszentrum Jülich GmbH (D-52425 Jülich, Germany). TGA/DTA measurements were carried out with a Mettler Toledo TGA/SDTA 851 in dry N₂ (60 mL min⁻¹) at a heating rate of 5 K min⁻¹. Vibrational spectra were recorded on a Bruker VERTEX 70 FT-IR spectrometer coupled with a RAM II FT-Raman module (1064 nm Nd:YAG laser) on KBr disks for the FT-IR and the solid material for the Diamond ATR-FTIR and Raman measurements. Liquid-phase ATR-FTIR spectra were obtained on saturated **Na-2** solution in H₂O. UV-Vis spectra were measured using 10 mm quartz cuvettes on Analytik Jena Specord S600 spectrophotometer.

**Synthesis of Na₂₅[Co₉(OH)₃(H₂O)₆(C₆H₅AsO₃)₂- (P₂W₁₅O₅₆)₃]·86H₂O (Na-1).** Na₁₂[α-P₂W₁₅O₅₆]·24H₂O (0.221 g, 0.05 mmol), CoCl₂ (0.023 g, 0.175 mmol), and C₆H₅AsO₃H₂ (0.010 g, 0.05 mmol) were dissolved in 6 mL of 0.66 M CH₃COONa buffer (pH 5.2) [28] under vigorous stirring in an oil bath at 60 °C for 4 days. The reaction mixture was then filtered, divided into several vials and left for evaporation at room temperature. Brown plate-like crystals of the side product Na₁₆₋ₓHₓ[(H₂O)₂Co₄(P₂W₁₅O₅₆)₂]·nH₂O (**Na-3**) start to appear immediately after cooling the reaction mixture to room temperature. They have to be removed by filtration, and filtration of **Na-3** has to be repeated in the next few days. Pure dark-pink crystals of **Na-1** form after 2 weeks. The crystals are collected by filtration, washed with 0.66 M CH₃COONa buffer and dried in air. Yield: 0.016 mg (6.6 % based on {P₂W₁₅}). Elemental analysis: calculated for C₁₂H₁₉₇Co₉Na₂₅O₂₆₉P₆As₂W₄₅ (found): C, 1.00 (1.04); H, 1.38 (1.38); Co, 3.69 (3.70); Na, 4.00 (4.00); P, 1.29 (1.29); As, 1.04 (1.03); W, 57.61 (55.7) %. IR (KBr pellet), cm⁻¹: 3454 (s, br); 1632 (s); 1439 (w); 1418 (w); 1385 (w); 1352 (w); 1084 (s);

1042 (m); 1011 (m); 933 (s); 912 (s); 880 (s); 806 (s); 731 (s); 598 (m); 561 (m); 525 (m); 492 (m); 459 (m);. Raman (in solid), cm⁻¹: 977 (s); 956 (s); 880 (m); 816 (m); 523 (w); 370 (m); 121 (m). UV-Vis ($H_2O$): $\lambda$ = 215 nm, $\varepsilon$ = 2.45×10⁷ M⁻¹cm⁻¹; $\lambda$ = 265 nm, $\varepsilon$ = 1.06×10⁷ M⁻¹cm⁻¹; $\lambda$ = 532 nm, $\varepsilon$ = 3.67×10⁵ M⁻¹cm⁻¹. UV-Vis (0.5 M $CH_3COONa$ solution, pH 5.1): $\lambda$ = 235 nm, $\varepsilon$ = 3.21×10⁹ M⁻¹cm⁻¹; $\lambda$ = 531 nm, $\varepsilon$ = 3.88×10⁵ M⁻¹cm⁻¹.

**Synthesis of $Na_{25}[Co_9(OH)_3(H_2O)_6(H_2NC_6H_4AsO_3)_2 (P_2W_{15}O_{56})_3]\cdot86 H_2O$ (Na-2).** $Na_{12}[\alpha\text{-}P_2W_{15}O_{56}]\cdot24 H_2O$ (0.221 g, 0.05 mmol), $CoCl_2$ (0.023 g, 0.175 mmol), and $H_2NC_6H_4AsO_3H_2$ (0.011 g, 0.05 mmol) were dissolved in 6 mL of 0.66 M $CH_3COONa$ buffer (pH 5.2) [28] under vigorous stirring in an oil bath at 60 °C for 4 days. Then the resulting reaction mixture was filtered, divided into several vials and left for evaporation at room temperature. Brown plate-like crystals of the side product $Na_{16-x}H_x[(H_2O)_2Co_4(P_2W_{15}O_{56})_2]\cdot nH_2O$ (Na-3) start to form immediately after cooling the reaction mixture and have to be removed by filtration after the reaction and on the next day. Pink needle-like crystals of Na-2 form after 2 days. They were collected by filtration, washed with 0.66 M $CH_3COONa$ buffer and dried in air. Yield: 0.0075 mg (3.1 % based on {$P_2W_{15}$}). Elemental analysis: calculated for $C_{12}H_{199}N_2Co_9Na_{25}O_{269}P_6As_2W_{45}$ (found): C, 1.00 (1.09); H, 1.39 (1.37); N, 0.19 (0.19); Co, 3.69 (3.73); Na, 3.99 (3.93); P, 1.29 (1.33); As, 1.04 (1.09); W, 57.49 (57.67)%. IR (KBr pellet), cm⁻¹: 3449 (s, br); 1626 (s); 1504 (w); 1470 (w); 1383 (w); 1354 (w); 1321 (w); 1292 (w); 1264 (w); 1194 (w); 1142 (m); 1086 (s); 1042 (m); 1009 (m); 932 (s); 910 (s); 881 (s); 806 (s); 725 (s); 600 (m); 561 (m); 521 (s); 457 (m). Raman (in solid), cm⁻¹: 978 (s); 957 (s); 916 (m), 905 (m), 893 (m), 883 (m); 839 (m); 523 (w). UV-Vis ($H_2O$): $\lambda$ = 195 nm, $\varepsilon$ = 1.23×10⁸ M⁻¹cm⁻¹; $\lambda$ = 255 nm, $\varepsilon$ = 3.01×10⁷ M⁻¹cm⁻¹; $\lambda$ = 532 nm, $\varepsilon$ = 6.60×10⁴ M⁻¹cm⁻¹.

**X-ray crystallography.** Single-crystal diffraction data for Na-1 and Na-2 were collected on a SuperNova (Agilent Technologies) diffractometer with MoKα radiation ($\lambda$ = 0.71073 Å) at 120 K. Crystals were mounted in a Hampton cryoloop with Paratone-N oil to prevent water loss. Absorption corrections were applied numerically based on multifaceted crystal model using CrysAlis software. [29] The SHELXTL software package [30] was used to solve and refine the structure. The structures were solved by direct methods and refined by full-matrix least-squares method against $|F|^2$ with anisotropic thermal parameters for all heavy atoms (W, P, Co, Na).

The hydrogen atoms of the phenyl rings in Na-1 and Na-2 and the amino groups in Na-2 were placed in geometrically calculated positions. The hydrogens of the water molecules and OH groups were not located. The relative site occupancy factors for the disordered positions of Na⁺ counterions and oxygen atoms of co-crystallized solvent water molecules were first refined in an isotropic approximation with $U_{iso}$= 0.05 and then fixed at the obtained values and refined without the thermal parameters restrictions.

One of the $P_2W_{15}$ ligands in Na-2 structure was disordered with 95:5 % ratio between two components. The disorder was modeled using combination of PART and EADP instructions. The rather high value still remaining for the second weighting term in (650 for Na-1 and 1420 for Na-2) most likely reflects the large volume occupied by highly disordered solvate and is consistent with large solvent-accessible volume remaining in the structures. Severe disorder did not allow localizing the positions for all the Na cations and O atoms of crystal waters in Na-1. Thus, only 16.5 Na⁺ ions and 60 co-crystallized $H_2O$ molecules have been found from the X-Ray data for this compound, while 25 Na⁺ and 86 crystal waters are present as based on elemental and thermogravimetrical analyses. For the overall consistency, the final formula in the CIF file of Na-1 correspond to the composition of the bulk materials determined by elemental analysis and TGA, as all further studies were performed on the isolated well-dried bulk materials of Na-1. The number of counterions and co-crystallized solvent molecules found in the crystal structure for Na-2 was consistent with those determined by other analytical techniques.

Additional crystallographic data are summarized in Table 2. Further details on the crystal structures investigation can be obtained, free of charge, on application to CCDC, 12 Union Road, Cambridge CB2 1EZ, UK: http://www.ccdc.cam.ac.uk/, e-mail: data_request@ccdc.cam.ac.uk, or fax: +441223 336033 upon quoting 1552113 (Na-1), and 1552114 (Na-2) numbers.

**Magnetic measurements** were performed using a Quantum Design MPMS-5XL SQUID magnetometer. The polycrystalline samples were compacted and immobilized into cylindrical PTFE capsules. In static magnetic field, data were acquired as a function of the magnetic field (0.1–5.0 T at 2.0 K) and temperature (2.0–290 K at 0.1 T). Dynamical field (ac) data were collected in the absence of a static bias field in the frequency range 1–1500 Hz ($T$ = 2.0–50 K, $B_{ac}$ = 3 G), however, no out-of-phase signals were detected. Data were corrected for the diamagnetic contributions of sample holder and compound (Na-1: $\chi_{dia}$ = −7.18×10⁻³ cm³ mol⁻¹, Na-2: $\chi_{dia}$ = −7.20×10⁻³ cm³ mol⁻¹).

**Table 2. Crystal data and structure refinement for Na-1 and Na-2**

| Sample | Na-1 | Na-2 |
|---|---|---|
| Empirical formula | $C_{12}H_{197}As_2Co_9Na_{25}O_{269}P_6W_{45}$ | $C_{12}H_{199}As_2Co_9N_2Na_{25}O_{269}P_6W_{45}$ |
| Formula weight, g mol⁻¹ | 14360.73 | 14390.76 |

| Crystal system | Triclinic | Triclinic |
|---|---|---|
| Space group | $P-1$ | $P-1$ |
| $a$, Å | 25.8088(3) | 14.2262(2) |
| $b$, Å | 25.8336(3) | 24.8597(4) |
| $c$, Å | 27.1598(3) | 37.9388(4) |
| $\alpha$ | 78.1282(11) ° | 81.9672(10)° |
| $\beta$ | 61.7276(14) ° | 87.8161(10)° |
| $\gamma$ | 60.6220(14) ° | 76.5409(12)° |
| Volume, Å³ | 13888.9(3) | 12920.6(3) |
| $Z$ | 2 | 2 |
| $D_{calc}$, g/cm³ | 3.434 | 3.699 |
| Absorption coefficient, mm⁻¹ | 19.496 | 20.958 |
| $F(000)$ | 12850 | 12882 |
| Crystal size, mm³ | 0.08× 0.23 × 0.81 | 0.11× 0.16 × 0.51 |
| Theta range for data collection | 4.08° – 25.35° | 4.08° – 25.68° |
| Completeness to $\Theta_{max}$ | 99.5% | 99.5% |
| Index ranges | $-31 \leq h \leq 30$, $-31 \leq k \leq 31$, $-32 \leq l \leq 32$ | $-17 \leq h \leq 17$, $-30 \leq k \leq 30$, $-45 \leq l \leq 46$ |
| Reflections collected | 259838 | 248406 |
| Independent reflections | 50595 | 48866 |
| $R_{int}$ | 0.0981 | 0.0799 |
| Observed ($I > 2\sigma(I)$) | 36674 | 41474 |
| Absorption correction | analytical using a multifaceted crystal model | |
| $T_{min} / T_{max}$ | 0.0201 / 0.3081 | 0.0159 / 0.2120 |
| Data / restraints / parameters | 50595 / 60 / 1941 | 48866 / 54 / 2189 |
| Goodness-of-fit on F² | 1.057 | 1.142 |
| $R_1$, $wR_2$ ($I > 2\sigma(I)$) | $R_1 = 0.0614$, $wR_2 = 0.1450$ | $R_1 = 0.0643$, $wR_2 = 0.1374$ |
| $R_1$, $wR_2$ (all data) | $R_1 = 0.0923$, $wR_2 = 0.1698$ | $R_1 = 0.0771$, $wR_2 = 0.1443$ |
| Largest diff. peak and hole, e Å⁻³ | 5.061 and −4.591 | 4.614 and −4.239 |

**Cyclic voltammograms of 1 – 4** in aqueous media were recorded using a Bio Logic SP-150 potentiostat controlled via EC-Lab software. The conventional three-electrode electrochemical cell included a glassy carbon working electrode with the diameter of 3 mm, a platinum wire counter electrode and an aqueous Ag/AgCl (3M NaCl) reference electrode (0.196 V $vs.$ SHE determined by measuring [Fe(CN)$_6$]$^{3-/4-}$ as an internal standard). The solu-tions were thoroughly deaerated with pure argon and kept under a positive Ar pressure during the experiments. Alumina powder was used for the cleaning of the working electrode which was then thoroughly rinsed with deionized water. Redox potentials were determined from the average values of the anodic and cathodic peak potentials and reported $vs.$ Ag/AgCl (3M NaCl) reference electrode.

**Template-stripped Au substrates**. Very flat Au$^{TS}$ surfaces were prepared according to the method reported by the Whiteside's group.[7] In brief, a 300–500 nm thick Au film is evaporated on a very flat silicon wafer covered by its native SiO$_2$ (RMS roughness of 0.4 nm) which was previously carefully cleaned by piranha solution (30 min in 2 : 1 H$_2$SO$_4$ : H$_2$O$_2$ (v/v). *Caution: Piranha solution is exothermic and strongly reacts with organics*), rinsed with deionized (DI) water and dried under a stream of nitrogen. A clean glass piece (ultrasonicated in acetone for 5 min, ultrasonicated in isopropanol for 5 min and UV-irradiated in ozone for 10 min) is glued (UV polymerizable glue) on the evaporated Au film and mechanically stripped with the Au film attached on the glass piece (Au film is cut with a razor blade around the glass piece). This very flat (RMS roughness of 0.4 nm same as the SiO$_2$ surface used as template) and clean template-stripped Au$^{TS}$ surface is immediately used for the formation of the molecule self-assembled monolayer.

**Spectroscopic ellipsometry**. We recorded spectroscopic ellipsometry data in the visible range using an UVISEL (Jobin Yvon Horiba) Spectroscopic Ellipsometer equipped with DeltaPsi 2 data analysis software. The system acquired a spectrum ranging from 2 to 4.5 eV (corresponding to 300 to 750 nm) with intervals of 0.1 eV (or 15 nm). Data were taken at an angle of incidence of 70°, and the compensator was set at 45.0°. Data were fitted by a regression analysis to a film-on-substrate model as described by their thickness and their complex refractive indexes. First, a background before monolayer deposition for the gold coated substrate was recorded. Secondly, after the monolayer deposition, we used a 2-layer model (substrate/SAM) to fit the measured data and to determine the SAM thickness. We employed the previously measured optical properties of the gold coated substrate (background), and we fixed the refractive index of the organic monolayer at 1.50. The usual values in the literature for the refractive index of organic monolayers are in the range 1.45–1.50.[3I] We can notice that a change from 1.50 to 1.55 would result in less than 1 Å error for a thickness less than 30 Å. We estimated the accuracy of the SAM thickness measurements at ± 2 Å.

**IRRAS measurements** were performed on a Bruker Vertex 70 FT-IR spectrometer equipped with a high-sensitivity Hg-Cd-Te (MCT) detector and an A513/Q variable angle reflection accessory including an automatic rotational holder for MIR polarizer. The IR beam was polarized with a KRS-5 polarizer with 99 % degree of polarization. Double-sided interferograms were collected with a sample frequency of 20 kHz, an aperture of 1.5 mm and a nominal spectral resolution of 4 cm$^{-1}$. The interferograms were apodized by a Blackmann-Harris 3-term apodization and zero-filled with a zero-filling factor of 2. The angle of incidence was set to 80°, and $p$-polarized IR radiation was used to record the spectra. For the background measurements, the sample chamber was purged with argon for 5 min, then 1024 scans were collected while continuing to purge. For the sample measurements, argon purging was started at the moment the first scan was recorded. The scans were averaged until the peaks arising from the water vapor in the sample chamber were compensated, for what typically 800–1500 scans were necessary. Where necessary, scattering correction was applied to the spectra.

*General procedure for the preparation of Au substrates for IRRAS*: Au substrates were fabricated by sputtering a 10 nm adhesive film of Ti and a 100 nm thick layer of Au on <100> oriented silicon wafers with a native SiO$_2$ layer. The freshly prepared Au substrates were cleaned in oxygen plasma [$p(O_2)$ = 0.4 mbar, $f$ = 40 kHz and $P$ = 75 W] for 4 min immediately prior to the **Na-2** deposition. For the deposition, a low concentrated solution (∼ 0.4 mmol) of the **Na-2** sample was prepared using ultrapure water with a conductivity of < 55 nS cm$^{-1}$. The Au substrate was stored for 24 h in this solution, then washed with a small amount of ultrapure water and dried for 24 h in a desiccator.

**Cyclic voltammograms of 2 on gold**. CV experiments were performed with a Modulab potentiostat from Solartron Analytical and a classical three-electrode electrochemical cell. The SAM covered Au$^{TS}$ electrode was used as the working electrode (WE). The counter electrode (CE) was a platinum wire (0.5 mm) and Ag/AgCl (saturated KCl) was used as a reference electrode (REF). CV curve was recorded at a scan rate 100 mV/s. The energy position (with respect to vacuum) of the LUMO of the SAM was estimated from the first reduction peak $E_{red}$ by $E_{LUMO}$ = − ($E_{red}$ + $E_{REF/SHE}$) − 4.24[32] in eV with $E_{REF/SHE}$ = 0.196 V for the Ag/AgCl reference electrode.

**XPS experiments of 2 on gold** were performed to analyze the chemical composition of the obtained layer on gold and to detect any unremoved contaminant. We used a Physical Electronics 5600 spectrometer fitted in an UHV chamber with a residual pressure of 2×10$^{-10}$ Torr. High resolution spectra were recorded with a monochromatic Al Kα X-ray source ($h\nu$ = 1486.6 eV), a detection angle of 45° as referenced to the sample surface, an analyzer entrance slit width of 400 μm and with an analyzer pass energy of 12 eV. In these conditions, the overall resolution as measured from the full-width half-maximum (FWHM) of the Ag 3d$_{5/2}$ line is 0.55 eV. The spectra were recalibrated with respect to the C 1s peak at 284.8 eV. Semi-quantitative analysis were completed after standard background subtraction according to Shirley's method.[33] Peaks were decomposed by using Voigt functions and a least-square minimization procedure and by keeping constant the Gaussian and Lorentzian broadenings for each component of a given peak.

**XPS experiments of the reference powder sample of Na-2** were carried out using the Specs Phoibos-150 energy analyzer and a non-monochromatized Al Kα X-ray source, with the overall energy resolution of 0.9 eV. The energy calibration of XPS measurements was done by aligning of C1s core level peak to 284.8 eV. The spectra were analyzed after Shirley background subtraction, with exceptions of Co2p, N1s and As3p+P2p regions, where linear background was used instead. Spectra were decomposed using Voigt-function shapes and fitting employed the KolXPD analysis package. Small charging effects were observed; C1s core-level exhibits an asymmetry and it was decomposed into two symmetrical Voigt components (287.1 and 288.5 eV). The energy axis of measurements was left unmodified.

**C-AFM measurements and TVS analysis**. We performed current-voltage measurements by C-AFM in ambient air (ICON, Bruker), using a PtIr-coated tip (tip radius of curvature less than 25 nm, force constant in the range 0.17–0.2 N/m). Placing the conducting tips at a stationary point contact formed nano-junctions. A square grid of 10×10 is defined with a lateral step of 2 nm. At each point, 10 *I-V* curves are acquired (back and forth) leading to the measurements of 1000 *I-V* traces. Out of these 1000 *I-V* traces, some were removed (main causes: no current - bad tip contact, noise larger than average current - tip contact fluctuations, too high current, e.g. short-circuit or pin-hole in the SAM, inducing saturation of the current preamplifier) leading to about 300 useful *I-V* traces (exact number indicated in the related figures). The load force was adjusted in the range 20–30 nN and measured by force-distance curves with the controlling software of the ICON. The bias was applied on the Au$^{TS}$ substrate and the tip was grounded through the input of the current amplifier. The voltage sweeps (back and forth) were applied from 0 to 1 V and then from 0 to –1 V.

The *I-V* curves are analyzed by the TVS (transient voltage spectroscopy) method.[26] In brief, the *I-V* data are plotted as $\ln(I/V^2)$ *vs.* $1/V$. A minimum in this curve corresponds to a transition from a direct tunneling electron transport through the molecules and a resonant tunneling via a frontier molecular orbital (LUMO or HOMO). The energy position $\varepsilon_0$ of the orbital involved in the transport mechanism with respect to the Fermi energy of the metal electrode is given by:

$$|\varepsilon_0| = 2 \frac{e|V_{T+}V_{T-}|}{\sqrt{V_{T+}^2 + 10|V_{T+}V_{T-}|/3 + V_{T-}^2}}$$

where $e$ is the electron charge, $V_{T+}$ and $V_{T-}$ are the voltage of the minima of the TVS plot at positive and negative voltages, respectively.[34]

CONCLUSIONS

In summary, we suggest a general strategy for functionalization of polyoxometalates by integration of organoarsonates as prosthetic co-ligands to polyoxotungstate units, which can also stabilize polynuclear magnetic cores. Depending on the terminal organoarsonate residue, this can enable their attachment to metallic electrode surfaces, which we intend to explore in the context of molecular electronics and spintronics. Two corresponding polyanions with external phenyl and *para*-aminophenyl groups have been prepared and characterized in the solid state and in solution. The derivative comprising terminal amino groups was anchored to an Au surface and the thus-obtained SAM was extensively characterized via ellipsometry, FT-IRRAS, XPS, cyclic voltammetry, and C-AFM measurements. This study thereby proves the recently suggested suitability of the –NH$_2$ functional group for direct binding of molecules to noble metal surfaces (e.g. Au$^0$) without a mercaptocarboxylate link commonly used for this purpose. Electron transport measurements by C-AFM show a relatively large dispersion of the current though the molecular junctions, corresponding to the polyoxotungstate-centered LUMO orbitals located between ca. 0.4 and 0.7 eV above the Fermi energy of the Au electrode. The presence of the second amino group not bound to the metal substrate in principle can also be used for post-functionalization of the formed SAM that we plan to explore in follow-up work.

ASSOCIATED CONTENT

**Supporting Information**. Bond valence sum values; (ATR)-IR, Raman, UV-Vis, XPS and IRRAS spectra; TGA curves, further CV details, AFM image, current histograms and TVS plots for SAM of **2** on Au surface, packing diagrams and crystallographic data for **Na-1** and **Na-2** in CIF format. This material is available free of charge via the Internet at http://pubs.acs.org

AUTHOR INFORMATION

**Corresponding Authors**

* paul.koegerler@ac.rwth-aachen.de
* dominique.vuillaume@iemn.fr

**Author Contributions**

The manuscript was written through contributions of all authors. All authors have given approval to the final version of the manuscript.

**Funding Sources**

EU ERC Starting Grant MOLSPINTRON, no. 308051; COST Action CM 1203 for STSM at IEMN

**Notes**

The authors declare no competing financial interests.

ACKNOWLEDGMENT

We gratefully acknowledge financial support by Forschungszentrum Jülich, EU ERC Starting Grant 1203 – MOLSPINTRON (P.K.) and COST Action CM 1203 (M. S.). We thank Brigitte Jansen for TGA measurements and Dr. Volkmar Heß for fruitful discussions.

REFERENCES

(1) *See for example*: (a) Sanvito, S.; Rocha, A. R. *J. Comput. Theor. Nanosci.* **2006**, *3*, 624-642; (b) Bogani, L.; Wernsdorfer, W. *Nat. Mater.* **2008**, *7*, 179-188; (c) Osorio, E. A.; Bjørnholm, T.; Lehn, J.-M.; Ruben, M.; van der Zant, H. S. J. *J. Phys. Condens. Matter* **2008**, *20*, 374121 / 1-14; (d) Sanvito, S. *Chem. Soc. Rev.* **2011**, *40*, 3336-3355; (e) Fahrendorf, S.; Atodiresei, N.; Besson, C.; Caciuc, V.; Matthes, F.; Blügel, S.; Kögerler, P.; Bürgler, D. E.; Schneider, C. M. *Nat. Commun.* **2013**, *4*, 2425/1-6; (f) Perrin, M. L.; Burzurí, E.; van der Zant, H. S. J. *Chem. Soc. Rev.* **2015**, *44*, 902-919, and references therein.

(2) *See for example*: (a) Lehmann, J.; Gaita-Ariño, A.; Coronado, E.; Loss, D. *Nat. Nanotechnol.* **2007**, *2*, 312-317; (b) Fang, X.; Kögerler, P. *Chem. Commun.* **2008**, 3396-3398; (c) Kortz, U.; Müller, A.; van Slageren, J.; Schnack, J.; Dalal, N. S.; Dressel, M. *Coord. Chem. Rev.* **2009**, *253*, 2315-2327; (d) Kögerler, P.; Tsukerblat, B.; Müller, A. *Dalton Trans.* **2010**, *39*, 21-36; (e) Clemente-Juan, J. M.; Coronado, E.; Gaita-Ariño, A. *Chem. Soc. Rev.* **2012**, *41*, 7464-7478; (f) Song, Y. F.; Tsunashima, R. *Chem. Soc. Rev.* **2012**, *41*, 7384-7402; (g) Izarova, N. V.; Kögerler P. in *Trends in Polyoxometalates Research*; Ruhlmann, L., Schaming, D., Eds.; Nova Science Publishers: Hauppauge, **2015**, 121-149; (h) Ji, Y. C.; Huang,

L. J.; Hu, J.; Streb, C.; Song, Y. F. *Energy Environ. Sci.* **2015**, *8*, 776–789; (i) Busche, C.; Vilà-Nadal, L.; Yan, J.; Miras, H. N.; Long, D.-L.; Georgiev, V. P.; Asenov, A.; Pedersen, R. H.; Gadegaard, N.; Mirza, M. M.; Paul, D. J.; Poblet, J. M.; Cronin, L. *Nature* **2014**, *515*, 545-549; (j) Palii, A.; Tsukerblat, B.; Clemente-Juan, J. M.; Coronado, E. *J. Phys. Chem. C* **2016**, *120*, 16994-17005; (k) Shiddiq, M.; Komijani, D.; Duan, Y.; Gaita-Ariño, A.; Coronado, E.; Hill, S. *Nature* **2016**, *531*, 348-351; (l) Linnenberg, O.; Moors, M.; Solé-Daura, A.; López, X.; Bäumer, C.; Kentzinger, E.; Pyckhout-Hintzen, W.; Monakhov, K. Yu. *J. Phys. Chem. C*, **2017**, *121*, 10419-10429.

(3) (a) Lissel, F.; Schwarz, F.; Blacque, O.; Riel, H.; Lörtscher, E.; Venkatesan, K.; Berke, H. *J. Am. Chem. Soc.* **2014**, *136*, 14560-14569; (b) Monakhov, K. Yu.; Moors, M.; Kögerler, P. *Adv. Inorg. Chem.* **2017**, *69*, 251–286, and references therein.

(4) *See for example*: (a) Dolbecq, A.; Dumas, E.; Mayer, C. R.; Mialane, P. *Chem. Rev.* **2010**, *110*, 6009-6048; (b) Proust, A.; Matt, B.; Villanneau, R.; Guillemot, G.; Gouzerh, P.; Izzet, G. *Chem. Soc. Rev.* **2012**, *41*, 7605-7622; (c) Santonia, M.-P.; Hanana, G. S.; Hasenknopf, B. *Coord. Chem. Rev.* **2014**, *281*, 64-85, and references therein.

(5) *See for example*: (a) El Moll, H.; Dolbecq, A.; Marrot, J.; Rousseau, G.; Haouas, M.; Taulelle, F.; Rogez, G.; Wernsdorfer, W.; Keita, B.; Mialane, P. *Chem. Eur. J.* **2012**, *18*, 3845-3849; (b) El Moll, H.; Zhu, W.; Oldfield, E.; Rodriguez-Albelo, M.; Mialane, P.; Marrot, J.; Vila, N.; Mbomekallé, I. M.; Rivière, E.; Duboc, C.; Dolbecq, A. *Inorg. Chem.* **2012**, *51*, 7921-7931; (c) Rousseau, G.; Rivière, E.; Dolbecq, A.; Marrot, J.; Oms, O.; Mialane, P. *Eur. J. Inorg. Chem.* **2013**, 1793-1798; (d) Saad, A.; Zhu, W.; Rousseau, G.; Mialane, P.; Marrot, J.; Haouas, M.; Taulelle, F.; Dessapt, R.; Serier-Brault, H.; Rivière, E.; Kubo, T.; Oldfield, E.; Dolbecq, A. *Chem. Eur. J.* **2015**, *21*, 1-12; (e) Saad, A.; Anwar, N.; Rousseau, G.; Mialane, P.; Marrot, J.; Haouas, M.; Taulelle, F.; Mc Cormac, T.; Dolbecq, A. *Eur. J. Inorg. Chem.* **2015**, 4775-4782; (f) Xue, H.; Zhao, J.-W.; Pan, R.; Yang, B.-F.; Yang, G.-Y.; Liu, H.-S. *Chem. Eur. J.* **2016**, *22*, 12322-12331; (g) Ban, R.; Sun, X.; Wang, J.; Ma, P.; Zhang, C.; Niu, J.; Wang, J. *Dalton Trans.* **2017**, *46*, 5856-5863.

(6) *See for example*: (a) Joo, N.; Renaudineau, S.; Delapierre, G.; Bidan, G.; Chamoreau, L. M.; Thouvenot, R.; Gouzerh, P.; Proust, A. *Chem. Eur. J.* **2010**, *16*, 5043-5051; (b) Musumeci, C.; Luzio, A.; Pradeep, C. P.; Miras, H. N.; Rosnes, M. H.; Song, Y.-F.; Long, D.-L.; Cronin, L.; Pignataro, B. *J. Phys. Chem. C* **2011**, *115*, 4446-4455; (c) Mercier, D.; Boujday, S.; Annabi, C.; Villanneau, R.; Pradier, C.-M.; Proust, A. *J. Phys. Chem. C* **2012**, *116*, 13217-13224; (d) Rinfray, C.; Izzet, G.; Pinson, J.; Gam Derouich, S.; Combellas, C.; Kanoufi, F.; Proust, A. *Chem. Eur. J.* **2013**, *19*, 13838-13846; (e) Derouich, S. G.; Rinfray, C.; Izzet, G.; Pinson, J.; Gallet, J.-J.; Kanoufi, F.; Proust, A.; Combellas, C. *Langmuir* **2014**, *30*, 2287-2296; (f) Yvon, C.; Surman, A. J.; Hutin, M.; Alex, J.; Smith, B. O.; Long, D.-L.; Cronin, L. *Angew. Chem. Int. Ed.* **2014**, *126*, 3404-3409; (g) Volatron, F.; Noël, J.-M.; Rinfray, C.; Decorse, P.; Combellas, C.; Kanoufi, F.; Proust, A. *J. Mater. Chem. C* **2015**, *3*, 6266-6275; (h) Lombana, A.; Rinfray, C.; Volatron, F.; Izzet, G.; Battaglini, N.; Alves, S.; Decorse, P.; Lang, P.; Proust, A. *J. Phys. Chem. C* **2016**, *120*, 2837-2845.

(7) *See for example*: (a) Weakley, T. J. R. *J. Chem. Soc. Chem. Commun.* **1984**, 1406-1407; (b) Gálán-Mascarós, J. R.; Gómez-García, C. J.; Borrás-Almenar, J.; Coronado, E. *Adv. Mater.* **1994**, *6*, 221-223; (c) Clemente-Juan, J. M.; Coronado, E.; Gálán-Mascarós, J. R.; Gómez-García, C. J. *Inorg. Chem.* **1999**, *38*, 55-63; (d) Ritchie, C.; Boyd, T.; Long, D.-L.; Ditzel, E.; Cronin, L. *Dalton Trans.* **2009**, 1587-1592; (e) Lydon, C.; Sabi, M. M.; Symes, M. D.; Long, D.-L.; Murrie, M.; Yoshii, S.; Nojirib, H.; Cronin, L. *Chem. Commun.* **2012**, *48*, 9819-9821.

(8) (a) Quek, S. Y.; Venkataraman, L.; Choi, H. J.; Louie, S. G.; Hybertsen, M. S.; Neaton, J. B. *Nano Lett.* **2007**, *7*, 3477-3482; (b) de la Llave, E.; Clarenc, R.; Schiffrin, D. J.; Williams, F. J. *J. Phys. Chem. C* **2014**, *118*, 468-475; (c) Koo, K. M.; Sina, A. A. I.; Carrascosa, L. G.; Shiddiky, M. J. A.; Trau, M. *Anal. Methods* **2015**, *7*, 7042-7054.

(9) Contant, R. *Inorg. Synth.* **1990**, *27*, 106-111.

(10) Finke R. G.; Droege, M. W. *Inorg. Chem.* **1983**, *22*, 1006-1008.

(11) (a) Goberna-Ferrón, S.; Vigara, L.; Soriano-López, J.; Gálán-Mascarós, J. R. *Inorg. Chem.* **2012**, *51*, 11707-11715; (b) Soriano-López, J.; Goberna-Ferrón, S.; Vigara, L.; Carbó, J. J.; Poblet, J. M.; Galán-Mascarós, J. R. *Inorg. Chem.* **2013**, *52*, 4753-4755; (c) Goberna-Ferrón, S.; Soriano-López, J.; Gálán-Mascarós, J. R.; Nyman, M. *Eur. J. Inorg. Chem.* **2015**, 2833-2840; (d) Yin, Q.; Tan, J. M.; Besson, C.; Geletii, Y. V.; Musaev, D. G.; Kuznetsov, A. E.; Luo, Z.; Hardcastle, K. I.; Hill, C. L. *Science* **2010**, *328*, 342-345; (e) Zhu, G.; Geletii, Y. V.; Kögerler, P.; Schilder, H.; Song, J.; Lense, S.; Zhao, C.; Hardcastle, K. I.; Musaev, D. G.; Hill, C. L. *Dalton Trans.* **2012**, *41*, 2084-2090; (f) Lv, H.; Song, J.; Geletii, Y. V.; Vickers, J. W.; Sumliner, J. M.; Musaev, D. G.; Kögerler, P.; Zhuk, P.; Bacsa, J.; Zhu, G.; Hill, C. L. *J. Am. Chem. Soc.* **2014**, *136*, 9268-9271; (g) Soriano-López, J.; Musaev, D. G.; Hill, C. L.; Galán-Mascarós, J. R.; Carbo, J. J.; Poblet, J. M. *J. Catal.* **2017**, *350*, 56-63; (h) Song, F.; Ding, Y.; Ma, B.; Wang, C.; Wang, Q.; Du, X.; Fua, S.; Song, J. *Energy Environ. Sci.* **2013**, *6*, 1170-1184.

(12) Unit cell for $Na_5H_{25-y}[Co_9(H_2O)_6(OH)_3(HPO_4)_2(\alpha-P_2W_{15}O_{56})_3] \cdot nH_2O$ (**Na-4**): Triclinic, *P*–1, *a* = 13.9486(4) Å, *b* = 29.2148(7) Å, *c* = 31.7155(8) Å, $\alpha$ = 74.603(2)°, $\beta$ = 81.760(2)°, $\gamma$ = 89.949(2)°, *V* = 12320.2(5) Å³, Z = 2.

(13) Barsukova, M.; Izarova, N. V.; Ngo Biboum, R.; Keita, B.; Nadjo, L.; Ramachandran, V.; Dalal, N. S.; Antonova, N. S.; Carbó, J. J.; Poblet, J. M.; Kortz, U. *Chem. Eur. J.* **2010**, *16*, 9076-9085.

(14) Lueken, H. *Magnetochemie*, Teubner, Stuttgart, **1999**.

(15) (a) Ballhausen, C. *Introduction to Ligand-Field Theory*, McGraw-Hill, New York, **1962**; (b) Figgs, N. B.; Hitchman, M. A. *Ligand-Field Theory and its Applications*, Wiley-VCH, New York, **2000**.

(16) *See for example*: (a) Ruhlmann, L.; Nadjo, L.; Canny, J.; Contant, R.; Thouvenot, R. *Eur. J. Inorg. Chem.* **2002**, 975-986; (b) Lisnard, L.; Mialane, P.; Dolbecq, A.; Marrot, J.; Clemente-Juan, J. M.; Coronado, E.; Keita, B.; de Oliveira, P.; Nadjo, L.; Sécheresse, F. *Chem. Eur. J.* **2007**, *13*, 3525-3536; (c) Ruhlmann, L.; Schaming, D.; Ahmed, I.; Courville, A.; Canny, J.; Thouvenot, R. *Inorg. Chem.* **2012**, *51*, 8202-8211; (d) Duan, Y.; Clemente-Juan, J. M.; Giménez-Saiz, C.; Coronado, E. *Inorg. Chem.* **2016**, *55*, 925-938.

(17) (a) Weiss, E.; Chiechi, R.; Kaufman, G.; Kriebel, J.; Li, Z.; Duati, M.; Rampi, M.; Whitesides, G. *J. Am. Chem. Soc.* **2007**, *129*, 4336-4349.

(18) Smekal, W.; Werner, W. S. M.; Powell, C. J. *Surf. Interf. Anal.* **2005**, *37*, 1059.

(19) Yeh J. J.; Lindau, I. *At. Data Nucl. Data Tables*, **1985**, *32*, 1-155.

(20) Tanuma, S.; Powell, C. J.; Penn, D. R. *Surf. Interf. Anal.* **1994**, *21*, 165.

(21) Allara, D. L.; Baca, A.; Pryde, C. A. *Macromolecules* **1978**, *11*, 1215-1220.

(22) (a) Engelkes, V. B.; Beebe, J. M.; Frisbie, C. D. *J. Phys. Chem. B* **2005**, *109*, 16801-16810; (b) Kim, T.-W.; Wang, G.; Lee, H.; Lee, T. *Nanotechnology* **2007**, *18*, 315204.

(23) Smaali, K.; Lenfant, S.; Karpe, S.; Oçafrain, M.; Blanchard, P.; Deresmes, D.; Godey, S.; Rochefort, A.; Roncali, J.; Vuillaume, D. *ACS Nano* **2010**, *4*, 2411-2421.

(24) Smaali, K.; Clément, N.; Patriarche, G.; Vuillaume, D. *ACS Nano* **2012**, *6*, 4639-4647.

(25) (a) Reuter, M. G.; Hersam, M. C.; Seideman, T.; Ratner, M. A. *Nano Lett.* **2012**, *12*, 2243-2248; (b) Trasobares, J.; Rech, J.; Jonckheere, T.; Martin, T.; Aleveque, O.; Levillain, E.; Diez-Cabanes, V.; Olivier, Y.; Cornil, J.; Nys, J. P.; Sivakumarasamy, R.; Smaali, K.; Leclere, P.; Fujiwara, A.; Théron, D.; Vuillaume, D.; Clément, N. *Nano Lett.* **2017**, *17*, 3215-3222.

(26) (a) Beebe, J. M.; Kim, B.; Gadzuk, J. W.; Frisbie, C. D.; Kushmerick, J. G. *Phys. Rev. Lett.* **2006**, *97*, 026801; (b) Riccœur, G.; Lenfant, S.; Guérin, D.; Vuillaume, D. *J. Phys. Chem. C* **2012**, *116*, 20722-20730.

(27) Contant, R. *Inorg. Synth.* **1990**, *27*, 105-106.

(28) The 0.66 M CH$_3$COONa buffer (pH 5.2) was prepared by dissolving 41 g of CH$_3$COONa and 9.14 mL of glacial CH$_3$COOH in 1000 mL of H$_2$O.

(29) CrysAlisPro, Agilent Technologies, 1.171.36.28 (release 01-02-2013 CrysAlis171 .NET).

(30) Sheldrick, G. M. *Acta Cryst.* **2008**, *A64*, 112-122.

(31) (a) Ulman, A. *An Introduction to Ultrathin Organic Films: From Langmuir-Blodgett to Self-assembly*, Academic Press, Boston, **1991**; (b) Parikh, A. N.; Allara, D. L.; Ben Azouz, I.; Rondelez, F. *J. Phys. Chem.* **1994**, *98*, 7577-7590.

(32) Cardona, C. M.; Li, W.; Kaifer, A. E.; Stockdale, D.; Bazan, G. C. *Adv. Mater.* **2011**, *23*, 2367-2371.

(33) Shirley, D. A. *Phys. Rev. B* **1972**, 5, 4709-4714.

(34) Bâldea, I. *Phys. Rev. B* **2012**, *85*, 035442.

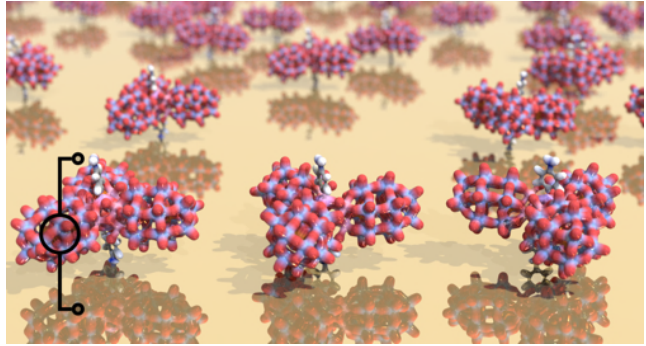

# Probing frontier orbital energies of {Co₉(P₂W₁₅)₃} polyoxometalate clusters at molecule–metal and molecule–water interfaces

Xiaofeng Yi,[†,‡] Natalya V. Izarova,[†] Maria Stuckart,[†,‡] David Guérin,[§] Louis Thomas,[§] Stéphane Lenfant,[§] Dominique Vuillaume,[§,*] Jan van Leusen,[‡] Tomáš Duchoň,[∥] Slavomír Nemšák,[†,¶] Svenja D. M. Bourone,[‡] Sebastian Schmitz[‡] and Paul Kögerler[†,‡,*]

[†]  Jülich-Aachen Research Alliance (JARA-FIT) and Peter Grünberg Institute 6, Forschungszentrum Jülich, D-52425 Jülich, Germany

[‡]  Institute of Inorganic Chemistry, RWTH Aachen University, D-52074 Aachen, Germany

[§]  Institute of Electronics, Microelectronics and Nanotechnology, CNRS, University of Lille, 59652 Villeneuve d'Ascq, France

[∥]  Faculty of Mathematics and Physics, Charles University, 18000 Prague, Czech Republic

[¶]  BESSY-II, Helmholtz Zentrum Berlin, D-12489 Berlin, Germany

# Content

# I. Bond valence sum calculations

**Table S1**. Bond valence sum values for different atoms in **Na-1**

| W, P, Co, and As centers | | W, P, Co, and As centers | | Terminal oxygens of W=O | |
|---|---|---|---|---|---|
| W1 | 6.32 | W41 | 6.26 | | |
| W2 | 6.35 | W42 | 6.22 | O17T | −1.77 |
| W3 | 6.28 | W43 | 6.03 | O18T | −1.77 |
| W4 | 6.27 | W44 | 6.17 | O19T | −1.69 |
| W5 | 6.29 | W45 | 6.21 | O20T | −1.71 |
| W6 | 6.27 | P1 | 4.71 | O21T | −1.90 |
| W7 | 6.23 | P2 | 4.73 | O22T | −1.84 |
| W8 | 6.31 | P3 | 4.63 | O23T | −1.91 |
| W9 | 6.34 | P4 | 4.85 | O24T | −1.66 |
| W10 | 6.30 | P5 | 4.72 | O25T | −1.60 |
| W11 | 6.33 | P6 | 4.73 | O26T | −1.81 |
| W12 | 6.11 | Co1 | 1.93 | O27T | −1.60 |
| W13 | 6.26 | Co2 | 2.00 | O28T | −1.96 |
| W14 | 6.22 | Co3 | 2.03 | O29T | −1.73 |
| W15 | 6.17 | Co4 | 1.99 | O30T | −1.76 |
| W16 | 6.24 | Co5 | 1.95 | O31T | −1.73 |
| W17 | 6.29 | Co6 | 2.04 | O32T | −1.92 |
| W18 | 6.14 | Co7 | 2.04 | O33T | −1.67 |
| W19 | 6.23 | Co8 | 1.97 | O34T | −1.76 |
| W20 | 6.22 | Co9 | 2.00 | O35T | −1.77 |
| W21 | 6.30 | As1 | 5.11 | O36T | −1.90 |
| W22 | 6.49 | As2 | 5.26 | O37T | −1.67 |
| W23 | 6.37 | Terminal oxygens of | | O38T | −1.76 |
| W24 | 6.13 | W=O | | O39T | −1.74 |
| W25 | 6.06 | O1T | −1.77 | O40T | −1.72 |
| W26 | 6.35 | O2T | −1.64 | O41T | −1.66 |
| W27 | 6.04 | O3T | −1.82 | O42T | −1.79 |
| W28 | 6.17 | O4T | −1.63 | O43T | −1.62 |
| W29 | 6.21 | O5T | −1.69 | O44T | −1.65 |
| W30 | 6.33 | O6T | −1.66 | O45T | −1.63 |
| W31 | 6.28 | O7T | −1.72 | Terminal aqua–ligands | |
| W32 | 6.10 | O8T | −1.91 | of Co−O$_{H2O}$ | |
| W33 | 6.28 | O9T | −1.77 | O2C | −0.36 |
| W34 | 6.23 | O10T | −1.97 | O3C | −0.30 |
| W35 | 6.13 | O11T | −1.71 | O4C | −0.30 |
| W36 | 6.23 | O12T | −1.66 | O5C | −0.36 |
| W37 | 6.34 | O13T | −1.87 | O7C | −0.35 |
| W38 | 6.24 | O14T | −1.63 | | |
| W39 | 6.37 | O15T | −1.66 | O9C | −0.35 |
| W40 | 6.18 | O16T | −1.73 | | |

**Table S1**. Bond valence sum values for different atoms in **Na-1** (continuation)

| μ₂-O (W–O–Co) | | μ₂-O (W–O–W) | | μ₂-O (W–O–W) | |
|---|---|---|---|---|---|
| O0C1 | −1.86 | O164 | −1.95 | O356 | −2.05 |
| O5C1 | −1.95 | O167 | −1.89 | O362 | −1.95 |
| O1C2 | −2.03 | O168 | −1.88 | O367 | −2.22 |
| O2C2 | −1.92 | O169 | −1.98 | O373 | −2.02 |
| O3C3 | −1.86 | O170 | −1.96 | O378 | −2.09 |
| O4C3 | −1.93 | O171 | −1.95 | O384 | −1.99 |
| O0C4 | −1.98 | O178 | −1.90 | O389 | −2.22 |
| O5C4 | −1.81 | O182 | −2.01 | O395 | −2.07 |
| O6C5 | −1.90 | O183 | −1.96 | O401 | −2.07 |
| O7C5 | −1.90 | O190 | −2.00 | O405 | −1.92 |
| O8C6 | −1.89 | O194 | −2.20 | O410 | −2.05 |
| O9C6 | −1.90 | O195 | −2.03 | O412 | −1.85 |
| O0C7 | −1.92 | O201 | −2.22 | O423 | −2.02 |
| O1C7 | −2.06 | O206 | −2.05 | O434 | −1.96 |
| O3C8 | −1.88 | O212 | −1.95 | O445 | −2.16 |
| O2C8 | −1.92 | O217 | −2.03 | O511 | −2.09 |
| O4C9 | −1.89 | O223 | −2.18 | O612 | −2.00 |
| O5C9 | −1.91 | O228 | −2.05 | O713 | −2.05 |
| μ₂-O (W–O–W) | | O234 | −1.99 | O814 | −2.05 |
| O12 | −2.01 | O239 | −2.09 | O915 | −2.09 |
| O13 | −1.95 | O240 | −2.07 | μ₃-O (2 W + P) | |
| O14 | −2.07 | O250 | −2.07 | O1P2 | −1.86 |
| O19 | −1.95 | O256 | −1.95 | O1P4 | −1.87 |
| O23 | −1.88 | O267 | −2.13 | O1P6 | −1.88 |
| O25 | −1.99 | O278 | −1.93 | O2P1 | −1.88 |
| O26 | −1.96 | O289 | −2.07 | O2P2 | −1.85 |
| O37 | −1.93 | O290 | −2.00 | O2P3 | −1.89 |
| O38 | −1.96 | O312 | −2.00 | O2P4 | −1.86 |
| O45 | −1.98 | O313 | −1.92 | O2P5 | −1.81 |
| O49 | −2.19 | O314 | −2.01 | O2P6 | −1.82 |
| O56 | −2.20 | O315 | −2.07 | O3P1 | −1.81 |
| O67 | −1.99 | O323 | −1.94 | O3P2 | −1.84 |
| O78 | −2.21 | O326 | −2.06 | O3P3 | −1.82 |
| O89 | −2.03 | O327 | −2.01 | O3P5 | −1.89 |
| O101 | −1.92 | O338 | −2.05 | O3P4 | −1.90 |
| O105 | −2.12 | O339 | −1.99 | O3P6 | −1.81 |
| O112 | −2.07 | O340 | −2.06 | O4P1 | −1.86 |
| O123 | −1.97 | O345 | −2.21 | O4P3 | −1.82 |
| O134 | −2.16 | O349 | −2.01 | O4P5 | −1.84 |
| O145 | −1.95 | O351 | −2.05 | | |

**Table S1**. Bond valence sum values for different atoms in **Na-1** (continuation)

| $\mu_3$-O (2 Co + As) | | $\mu_3$-O (3 Co) | | $\mu_4$-O (3 W + P) | |
|---|---|---|---|---|---|
| O1A1 | −2.15 | O1 | −1.00 | O1P1 | −1.91 |
| O1A2 | −2.03 | O2 | −1.06 | | |
| O2A1 | −2.06 | O3 | −1.05 | O1P3 | −1.88 |
| O2A2 | −1.99 | $\mu_4$-O (3 Co + P) | | | |
| O3A1 | −2.00 | O6P4 | −1.83 | O1P5 | −1.89 |
| O3A2 | −2.06 | O4P4 | −1.83 | | |
| | | O4P2 | −1.82 | | |

**Table S2** Bond valence sum values for different atoms in **Na-2**

| W, P, Co, and As centers | | W, P, Co, and As centers | | Terminal oxygens of W=O | |
|---|---|---|---|---|---|
| W1 | 6.05 | W42 | 6.08 | | |
| W2 | 6.12 | W43 | 6.13 | O17T | −1.70 |
| W3 | 6.16 | W44 | 6.09 | O18T | −1.72 |
| W4 | 6.19 | W45 | 6.32 | O19T | −1.66 |
| W5 | 6.29 | W42 | 6.08 | O20T | −1.64 |
| W6 | 6.16 | P1 | 4.79 | O21T | −1.57 |
| W7 | 6.20 | P2 | 4.66 | O22T | −1.95 |
| W8 | 6.31 | P3 | 4.74 | O23T | −1.84 |
| W9 | 6.15 | P4 | 4.60 | O24T | −1.76 |
| W10 | 6.29 | P5 | 4.78 | O25T | −1.61 |
| W11 | 6.30 | P6 | 4.56 | O26T | −1.84 |
| W12 | 6.19 | Co1 | 1.95 | O27T | −1.77 |
| W13 | 6.26 | Co2 | 2.07 | O28T | −1.98 |
| W14 | 6.18 | Co3 | 1.96 | O29T | −1.66 |
| W15 | 6.18 | Co4 | 1.93 | O30T | −1.64 |
| W16 | 6.26 | Co5 | 1.96 | O31T | −1.75 |
| W17 | 6.20 | Co6 | 2.01 | O32T | −1.62 |
| W18 | 6.22 | Co7 | 1.97 | O33T | −1.94 |
| W19 | 6.25 | Co8 | 2.02 | O34T | −1.73 |
| W20 | 6.26 | Co9 | 2.05 | O35T | −2.06 |
| W21 | 6.19 | As1 | 4.93 | O36T | −1.74 |
| W22 | 6.24 | As2 | 4.99 | O37T | −1.66 |
| W23 | 6.32 | Terminal oxygens of W=O | | O38T | −1.89 |
| W24 | 6.32 | | | O39T | −1.74 |
| W25 | 6.19 | O1T | −1.87 | O40T | −1.87 |
| W26 | 6.17 | O2T | −1.89 | O41T | −1.59 |
| W27 | 6.20 | O3T | −1.97 | O42T | −1.86 |
| W28 | 6.36 | O4T | −1.69 | O43T | −1.84 |
| W29 | 6.06 | O5T | −1.79 | O44T | −1.57 |
| W30 | 6.23 | O6T | −1.78 | O45T | −1.72 |
| W31 | 6.12 | O7T | −1.96 | Terminal aqua–ligands of Co–$O_{H2O}$ | |
| W32 | 6.09 | O8T | −1.69 | | |
| W33 | 6.31 | O9T | −1.70 | O1CH | −0.36 |
| W34 | 6.15 | O10T | −1.67 | O3CH | −0.33 |
| W35 | 6.29 | O11T | −1.77 | O5CH | −0.31 |
| W36 | 6.42 | O12T | −1.65 | O6CH | −0.33 |
| W37 | 6.16 | O13T | −1.62 | O8CH | −0.35 |
| W38 | 6.31 | O14T | −1.72 | | |
| W39 | 6.38 | O15T | −1.77 | O9CH | −0.33 |
| W40 | 6.12 | O16T | −1.63 | | |

**Table S2** Bond valence sum values for different atoms in **Na-2** (continuation)

| μ₂-O (W–O–Co) | | μ₂-O (W–O–W) | | μ₂-O (W–O–W) | |
|---|---|---|---|---|---|
| O0C1 | −1.92 | O123 | −2.13 | O356 | −2.21 |
| O1C1 | −1.86 | O134 | −2.01 | O362 | −2.06 |
| O2C2 | −1.96 | O145 | −2.10 | O367 | −1.97 |
| O3C2 | −1.90 | O164 | −1.95 | O373 | −2.06 |
| O4C3 | −1.85 | O169 | −1.99 | O378 | −2.22 |
| O5C3 | −1.92 | O170 | −1.87 | O384 | −2.01 |
| O5C4 | −2.01 | O171 | −2.57 | O389 | −2.04 |
| O0C4 | −2.06 | O182 | −2.55 | O395 | −2.07 |
| O6C5 | −2.21 | O183 | −1.91 | O401 | −1.96 |
| O7C5 | −2.13 | O190 | −1.97 | O405 | −2.09 |
| O8C6 | −2.15 | O194 | −2.16 | O410 | −2.00 |
| O9C6 | −2.17 | O195 | −2.23 | O412 | −2.04 |
| O5C7 | −1.89 | O201 | −2.27 | O423 | −1.93 |
| O0C7 | −1.82 | O206 | −2.15 | O434 | −2.07 |
| O2C8 | −1.84 | O212 | −1.93 | O445 | −2.05 |
| O1C8 | −1.93 | O217 | −2.47 | O511 | −2.08 |
| O3C9 | −1.93 | O223 | −2.22 | O612 | −1.98 |
| O4C9 | −1.90 | O228 | −2.47 | O713 | −2.02 |
| μ₂-O (W–O–W) | | O234 | −1.99 | O814 | −2.01 |
| O167 | −1.88 | O239 | −2.29 | O915 | −1.99 |
| O168 | −1.81 | O240 | −2.20 | μ₃-O (2 W + P) | |
| O178 | −1.80 | O250 | −2.07 | O1P2 | −1.85 |
| O12 | −1.88 | O256 | −1.95 | O1P4 | −1.85 |
| O13 | −1.90 | O267 | −2.07 | O1P6 | −1.83 |
| O14 | −2.05 | O278 | −2.01 | O2P1 | −1.82 |
| O15 | −1.96 | O289 | −2.03 | O2P2 | −1.83 |
| O23 | −1.99 | O290 | −1.88 | O2P3 | −1.86 |
| O26 | −1.95 | O312 | −1.95 | O2P4 | −1.80 |
| O27 | −2.00 | O313 | −1.93 | O2P5 | −1.86 |
| O38 | −2.04 | O314 | −2.02 | O2P6 | −1.81 |
| O39 | −2.04 | O319 | −2.02 | O3P1 | −1.87 |
| O45 | −2.17 | O323 | −1.90 | O3P2 | −1.79 |
| O49 | −2.04 | O325 | −2.00 | O3P3 | −1.85 |
| O56 | −2.13 | O326 | −1.93 | O3P4 | −1.83 |
| O67 | −2.19 | O337 | −1.97 | O3P5 | −1.86 |
| O78 | −2.05 | O338 | −1.89 | O3P6 | −1.83 |
| O89 | −2.21 | O340 | −2.02 | O4P1 | −1.84 |
| O101 | −2.11 | O345 | −2.17 | O4P3 | −1.86 |
| O105 | −1.94 | O349 | −2.15 | O4P5 | −1.84 |
| O112 | −2.00 | O351 | −2.01 | | |

**Table S2** Bond valence sum values for different atoms in **Na-2** (continuation)

| $\mu_3$-O (2 Co + As) | | $\mu_3$-O (3 Co) | | $\mu_4$-O (3 W + P) | |
|---|---|---|---|---|---|
| O1A1 | −2.02 | O1 | −1.05 | O1P1 | −1.91 |
| O2A1 | −2.08 | O2 | −1.03 | | |
| O3A1 | −1.98 | O3 | −1.03 | O1P3 | −1.99 |
| O1A2 | −2.10 | $\mu_4$-O (3 Co + P) | | | |
| O3A2 | −2.00 | O4P2 | −1.82 | O1P5 | −1.91 |
| O2A2 | −2.08 | O4P4 | −1.76 | | |
| | | O4P6 | −1.78 | | |

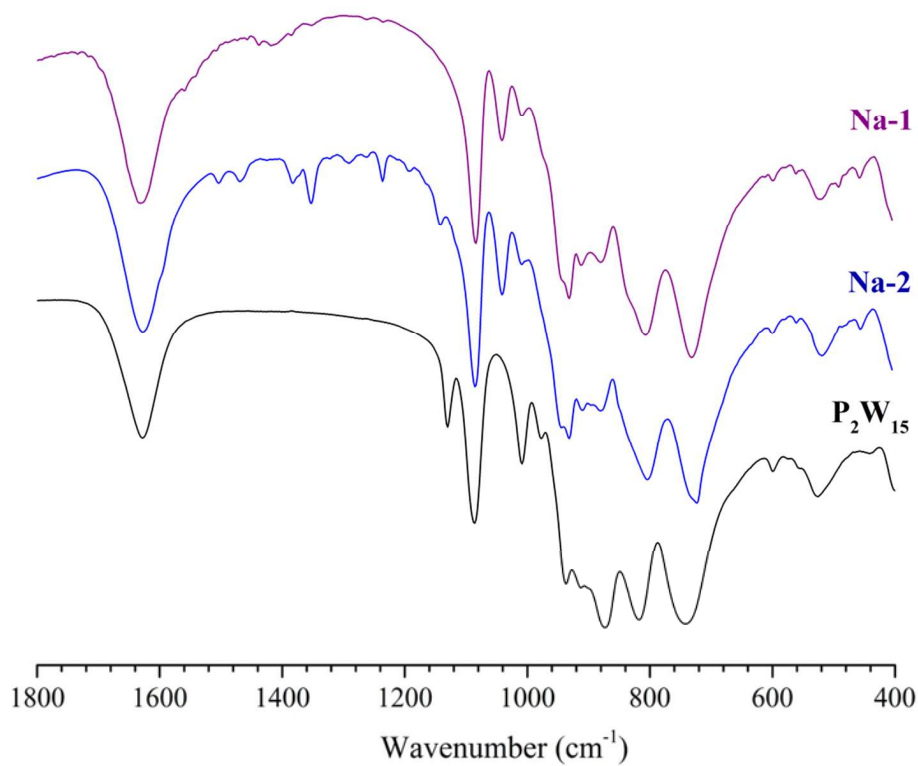

**Figure S1**. FT-IR spectra of **Na-1** (purple) and **Na-2** (blue) in comparison to that of the Na$_{12}$[$\alpha$-P$_2$W$_{15}$O$_{56}$]·24H$_2$O precursor (black).

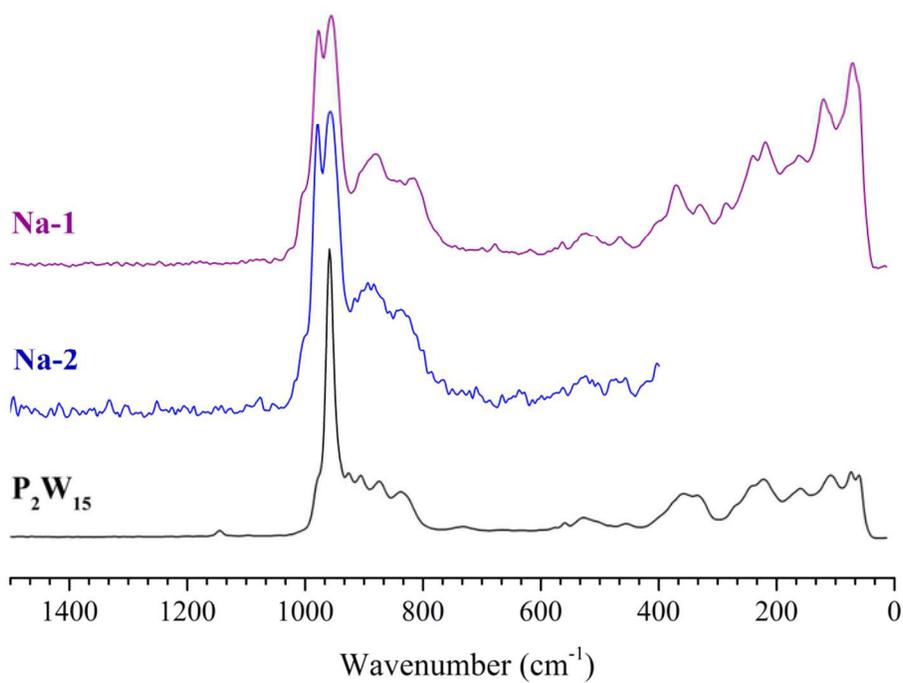

**Figure S2**. Raman spectra of **Na-1** (purple) and **Na-2** (blue) in comparison to that of the Na$_{12}$[$\alpha$-P$_2$W$_{15}$O$_{56}$]·24H$_2$O precursor (black).

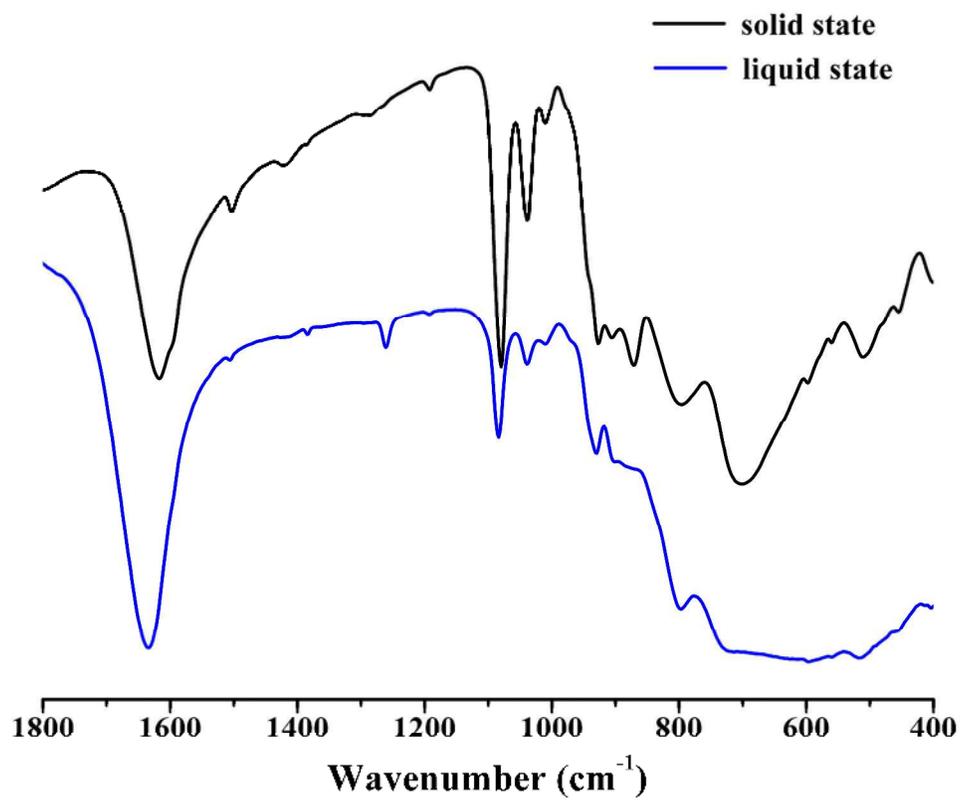

**Figure S3** ATR-FTIR spectra of **Na-2** in the solid state (black) and saturated aqueous solution (blue).

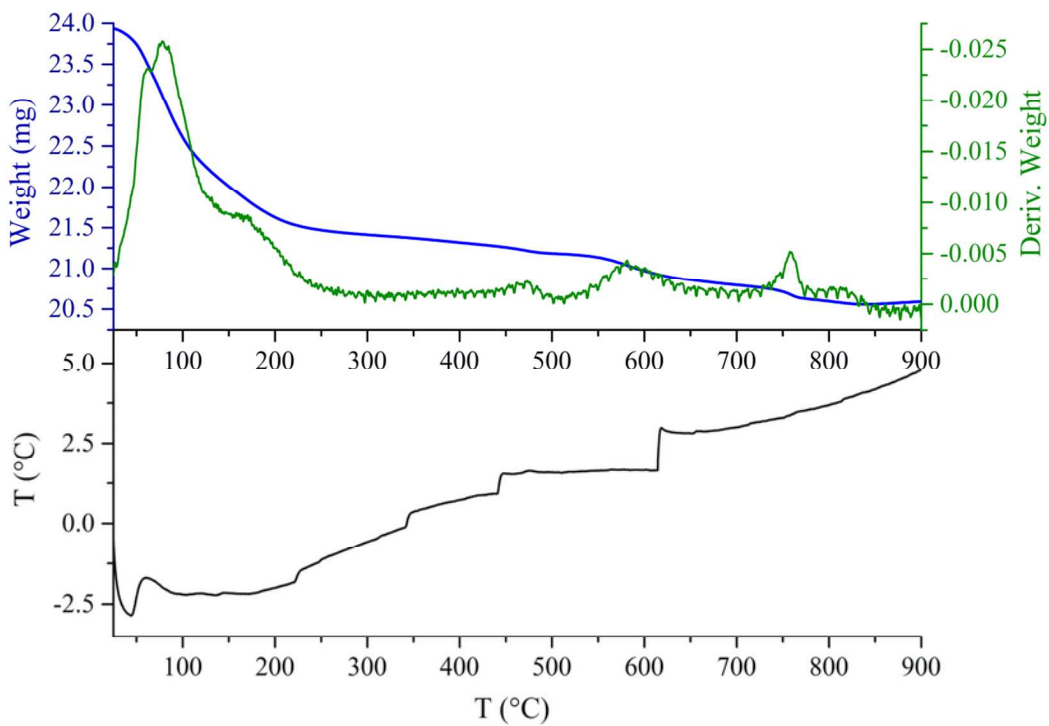

**Figure S4**. TGA (blue), differential TGA (green) and DTA (black) curves for **Na-1** from room temperature to 900 °C under $N_2$ atmosphere.

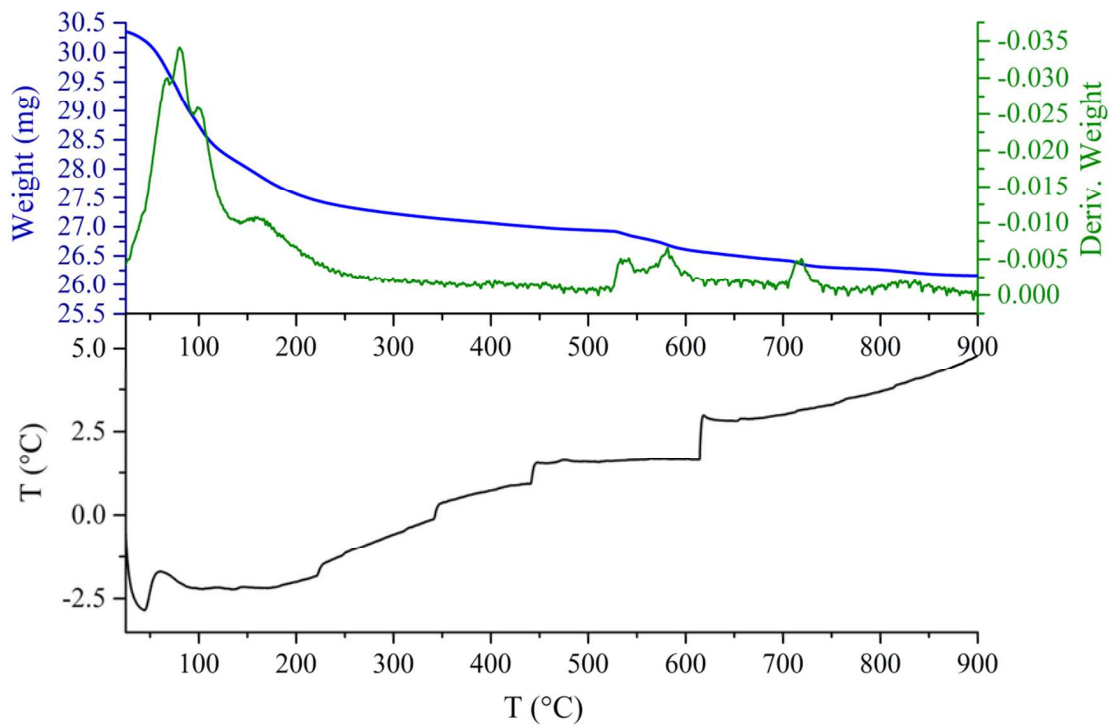

**Figure S5**. TGA (blue), differential TGA (green) and DTA (black) curves for **Na-2** from room temperature to 900 °C under $N_2$ atmosphere.

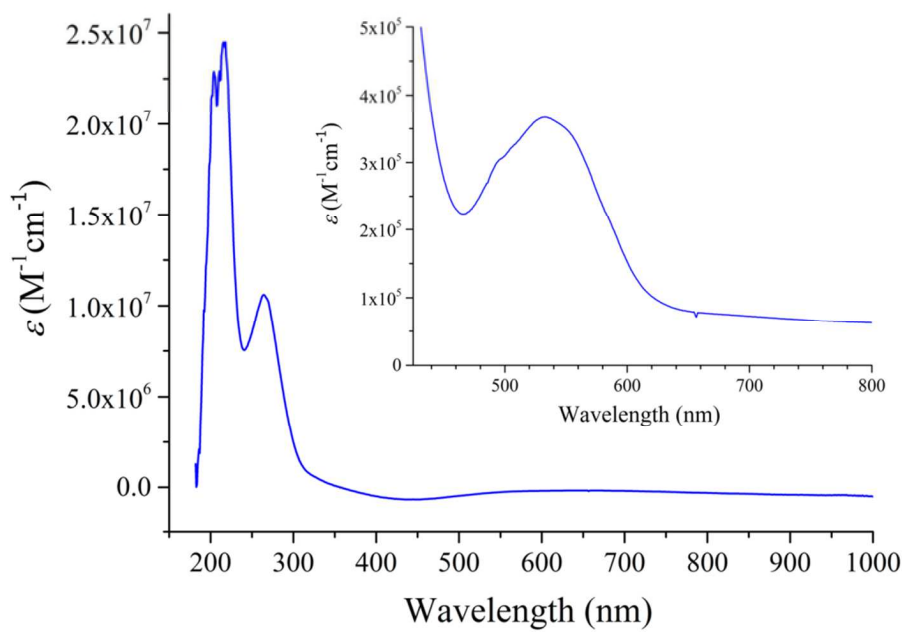

**Figure S6**. Room-temperature UV-Vis spectrum of **Na-1** solution in H$_2$O ($\varepsilon$ values are averaged from the spectra of the solutions with concentrations between 6.4·10$^{-6}$ M and 9.6·10$^{-6}$ M for the UV region and 2.8·10$^{-4}$ M for the visible light region).

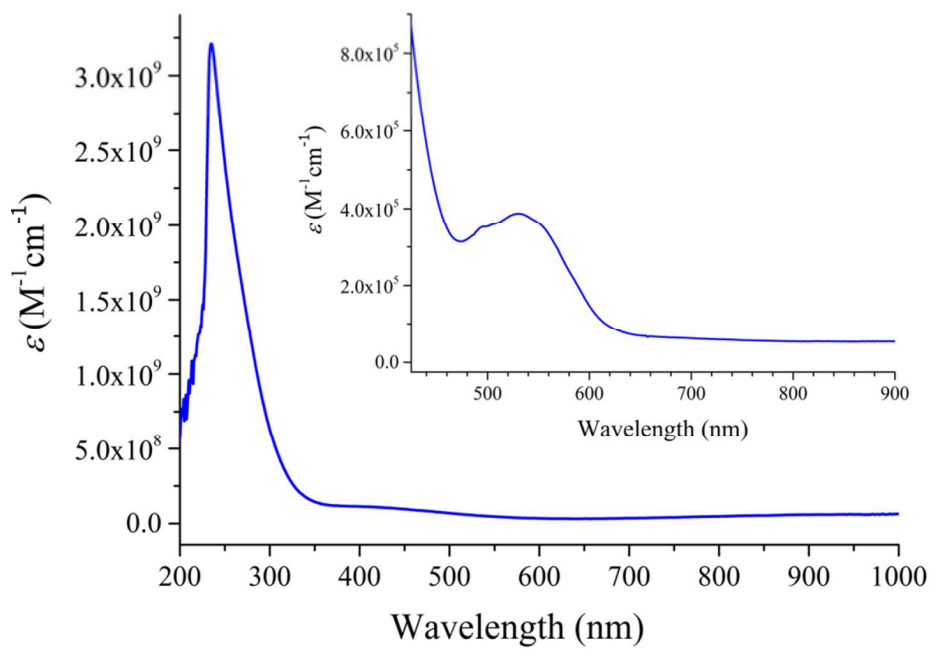

**Figure S7**. Room-temperature UV-Vis spectrum of **Na-1** solution in 0.5 M NaCH$_3$COO aqueous medium at pH 5.1 ($\varepsilon$ values are averaged from the spectra of the solutions with concentrations between 2.3·10$^{-8}$ M and 5.4·10$^{-7}$ M for the UV region and between 6.7·10$^{-5}$ M and 2.8·10$^{-4}$ M for the visible light region).

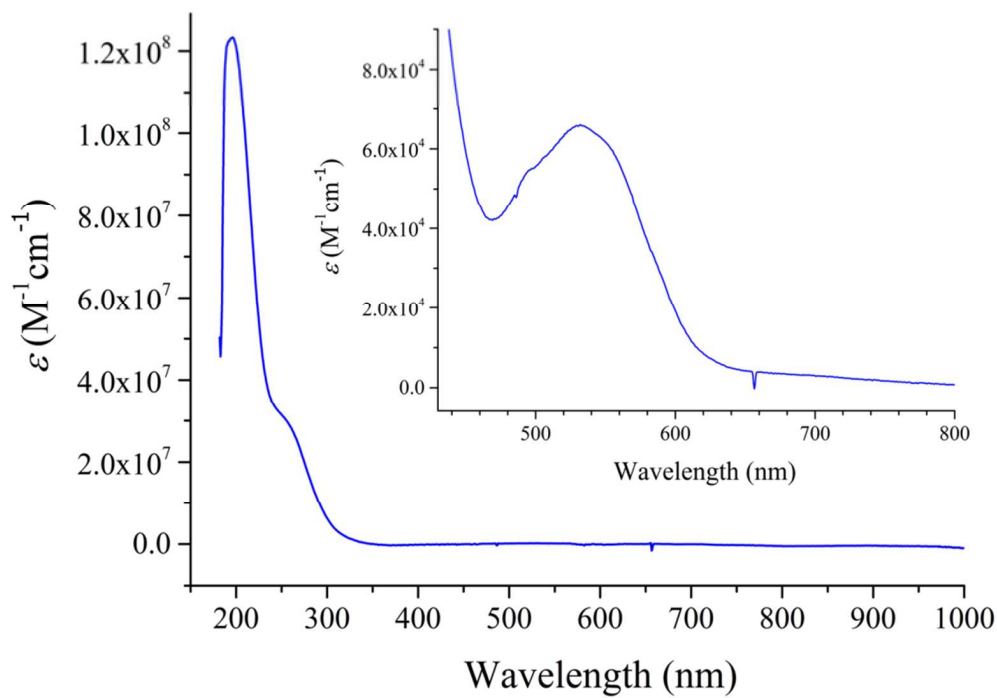

**Figure S8**. Room-temperature UV-Vis spectrum of **Na-2** solution in $H_2O$ ($\varepsilon$ values are averaged from the spectra of the solutions with concentrations between $1.1 \cdot 10^{-6}$ M and $2.2 \cdot 10^{-6}$ M).

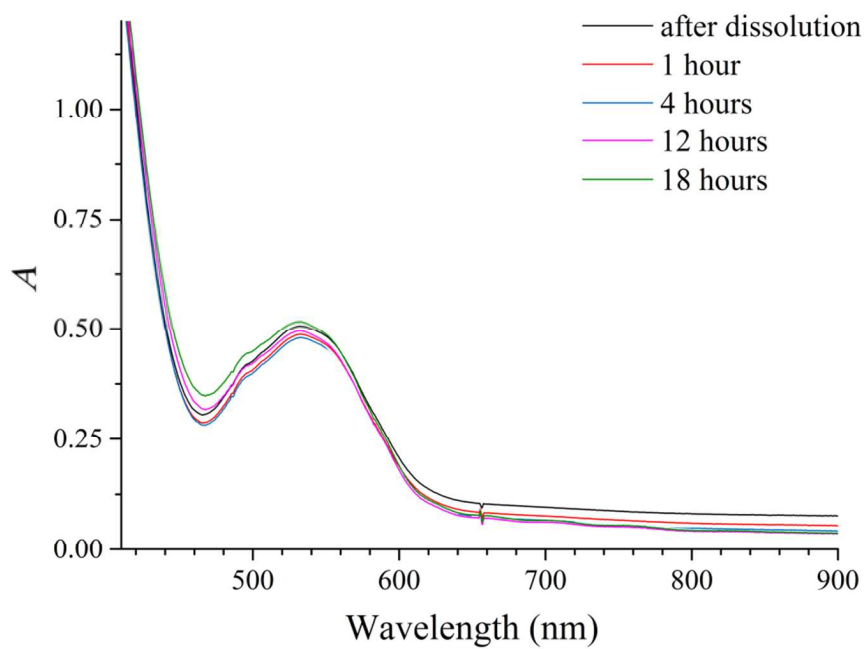

**Figure S9**. Time-dependent room-temperature UV-Vis spectrum of a $2.8 \cdot 10^{-4}$ M **Na-1** solution in $H_2O$.

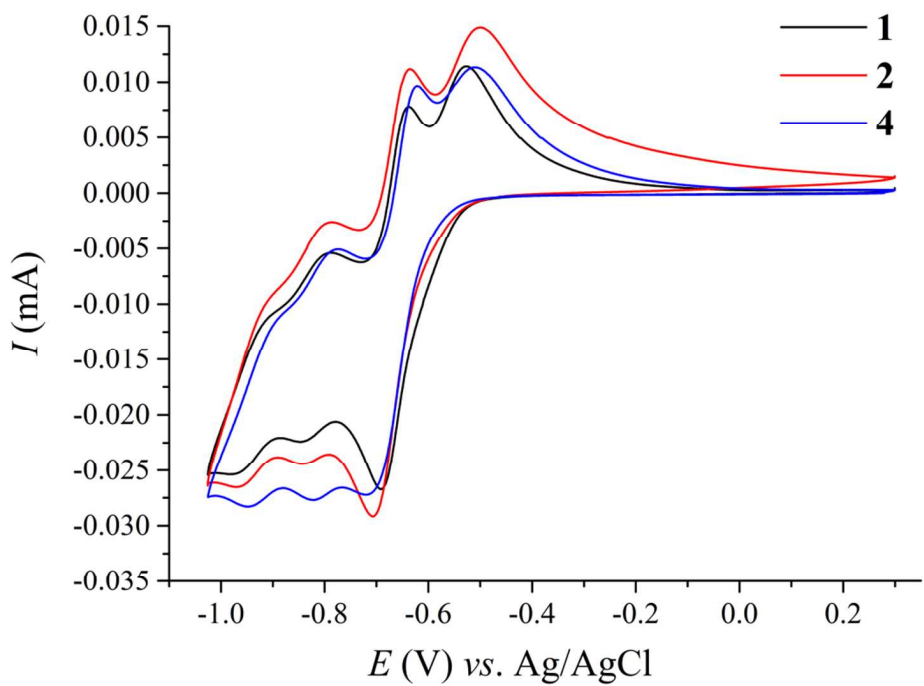

**Figure S10**. Comparison of cyclic voltammograms of 0.7 mM **Na-1** (black), **Na-2** (red) and **Na-4** (blue) solutions in 0.5 M $CH_3COONa$ buffer (pH 4.8), scan rate 20 mV / s.

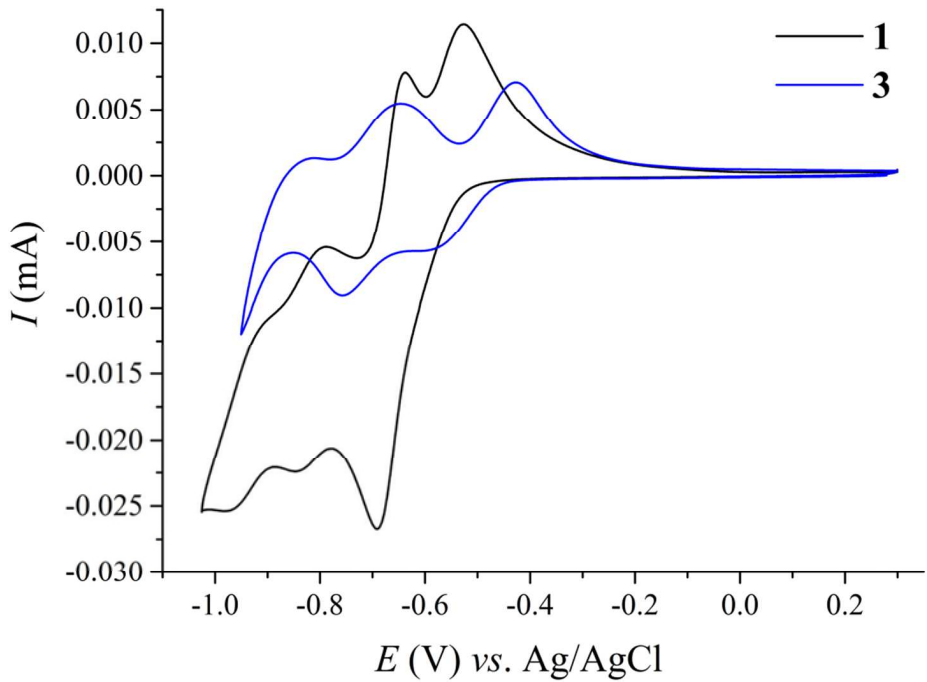

**Figure S11**. Comparison of cyclic voltammograms of **Na-1** (black) and **Na-3** (blue) 0.7 – 1 mM solutions in 0.5 M $CH_3COONa$ buffer (pH 4.8), scan rate 20 mV / s.

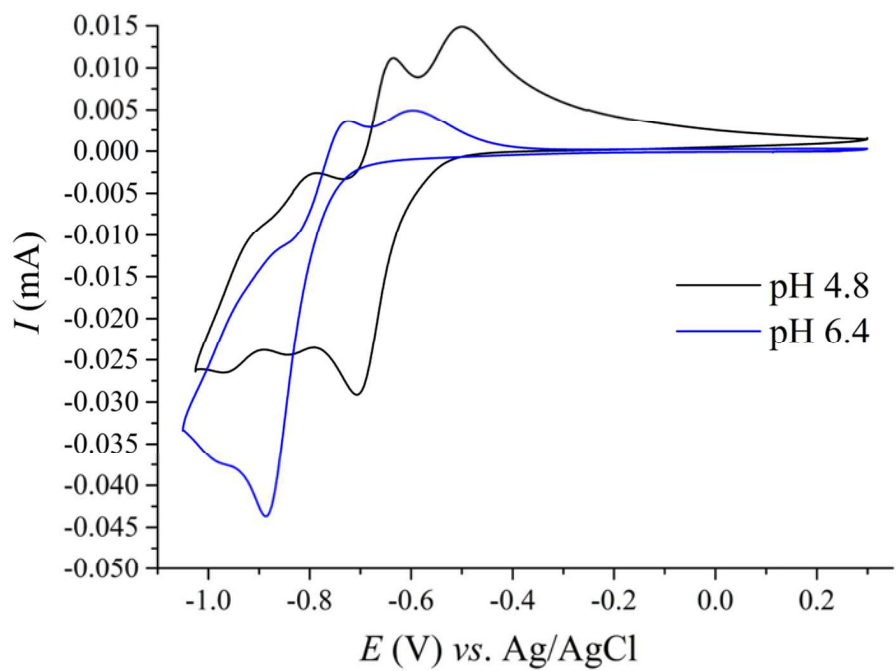

**Figure S12**. Cyclic voltammograms of 0.7 mM **Na-2** solutions in aqueous 0.5 M CH$_3$COONa media with pH 4.8 (black) and pH 6.4 (blue); scan rate 20 mV / s.

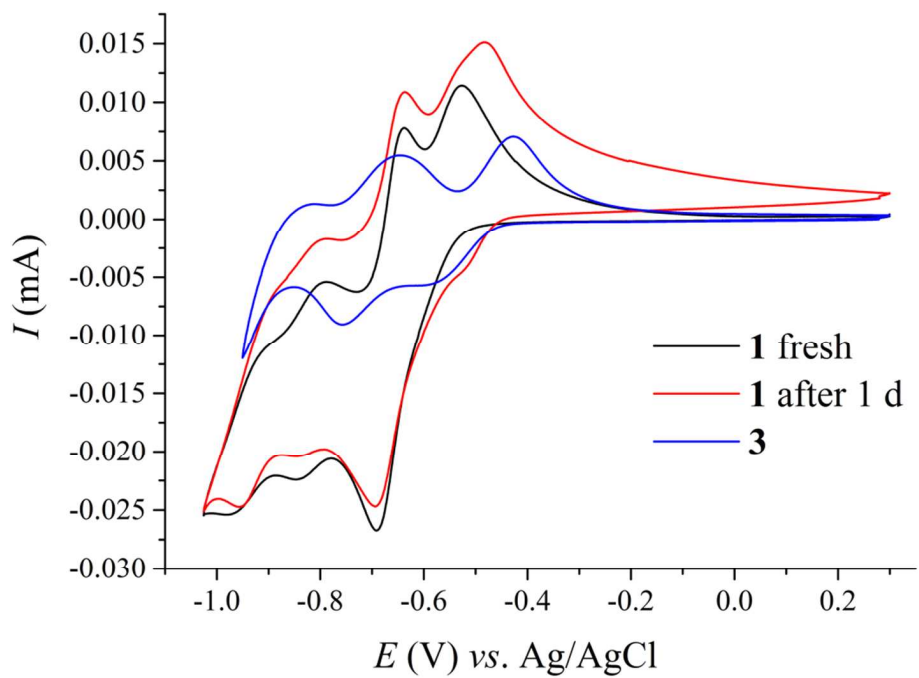

**Figure S13**. Cyclic voltammograms of 0.7 mM **Na-1** solutions in 0.5 M CH$_3$COONa buffer with pH 4.8: freshly prepared solution (black) and the solution after one day (red). A comparison with the CV of **Na-3** solution in the same medium.

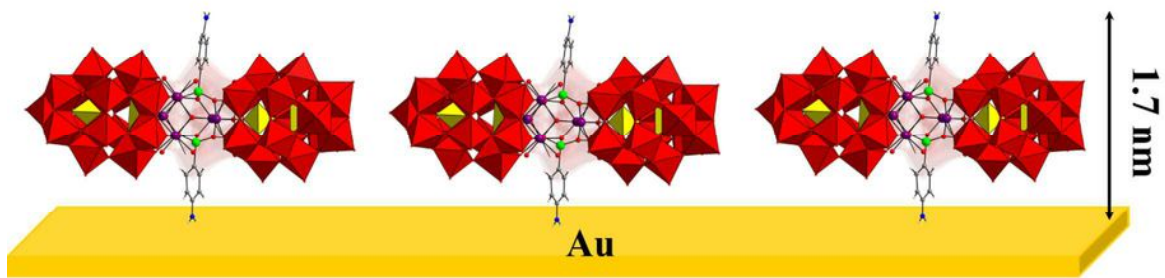

**Figure S14**. Representation of possible orientation of **2** on the gold surface.

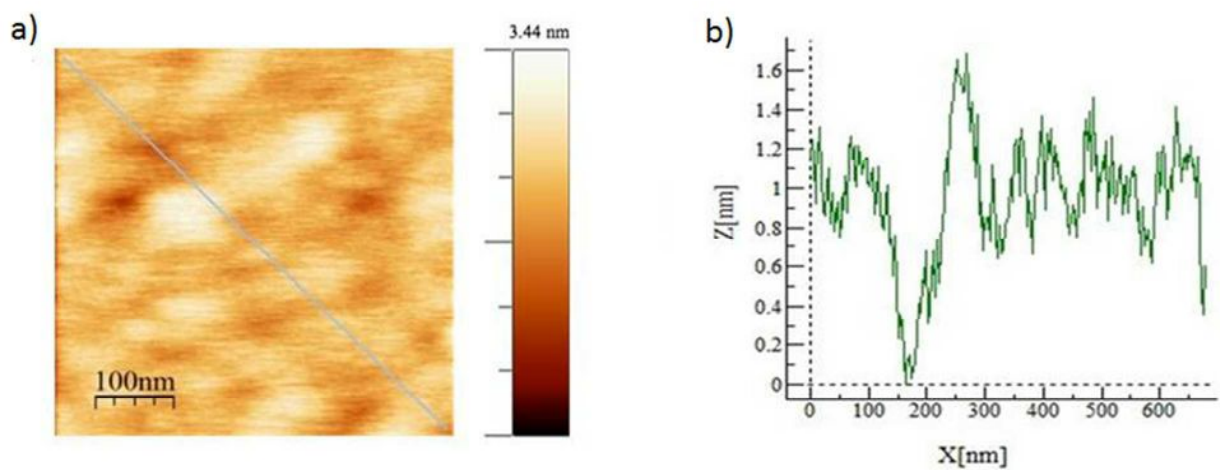

**Figure S15**. AFM image (**a**) and height profile corresponding to the blue line in the Fig. S15a (**b**) of **Na-2** on the gold surface.

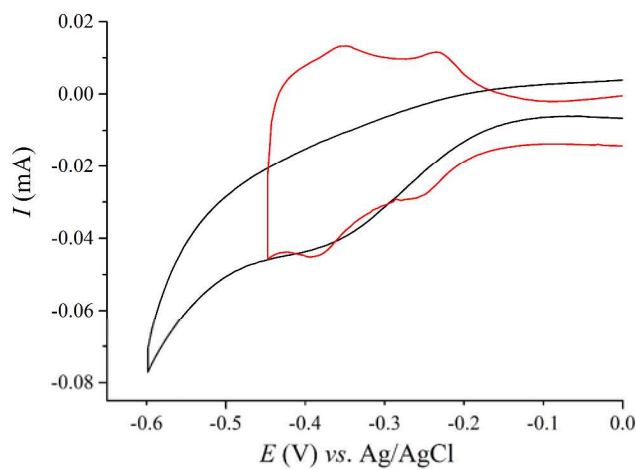

**Figure S16**. CV curves of SAM of **2** on gold surface (black line) with 0.66 M NaOAc/HOAc solution with pH 5.2 solution as an electrolyte and **Na-2** solution in the same medium at 100 mV/s.

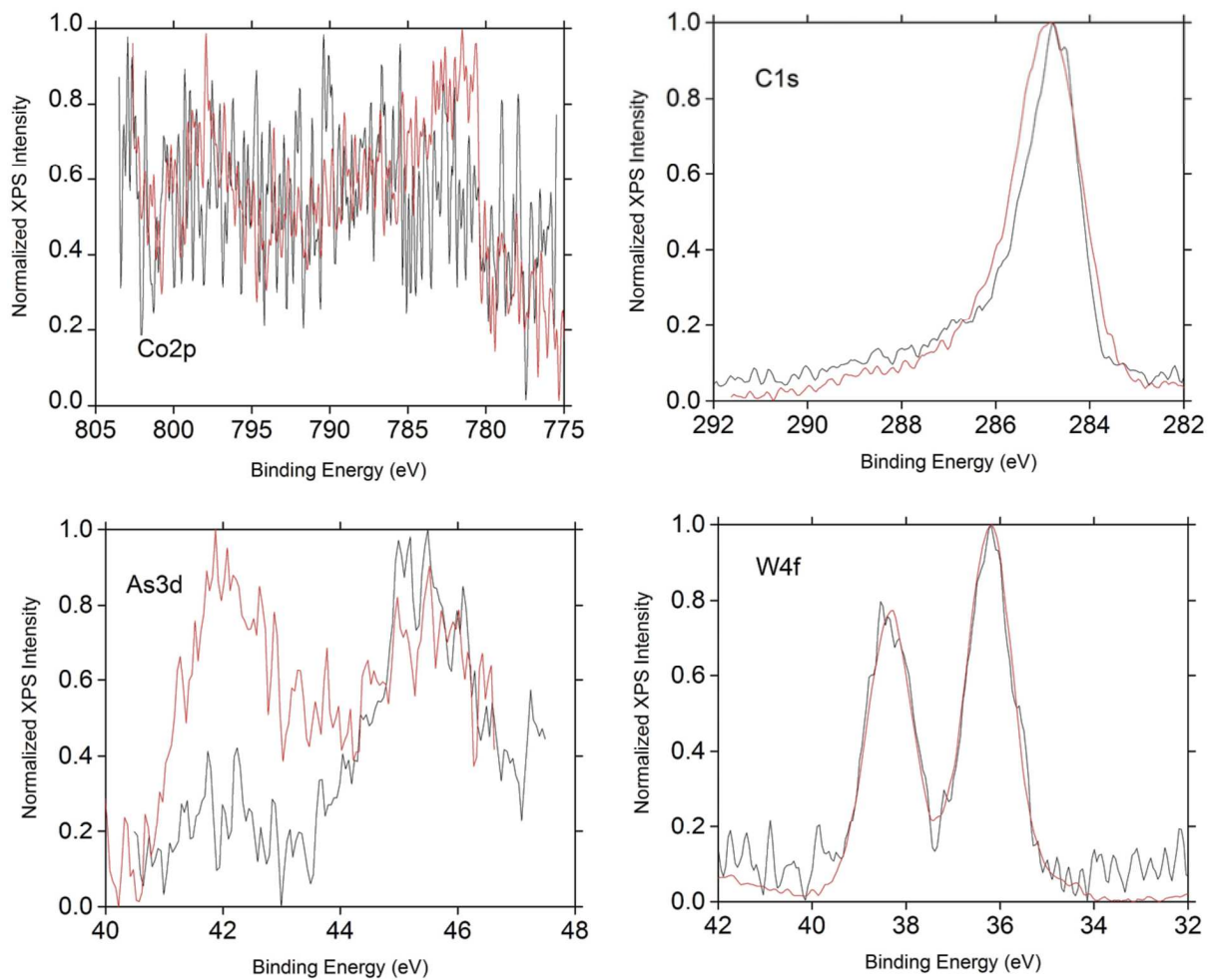

**Figure S17**. XPS spectra of **Na-2** SAM on a template stripped gold surface (black line; $10^{-3}$ M in DI water, 24h) and of a film of **Na-2** crystals on Au (red line; a drop of a concentrated solution of **Na-2** in deionized water was evaporated on a gold surface).

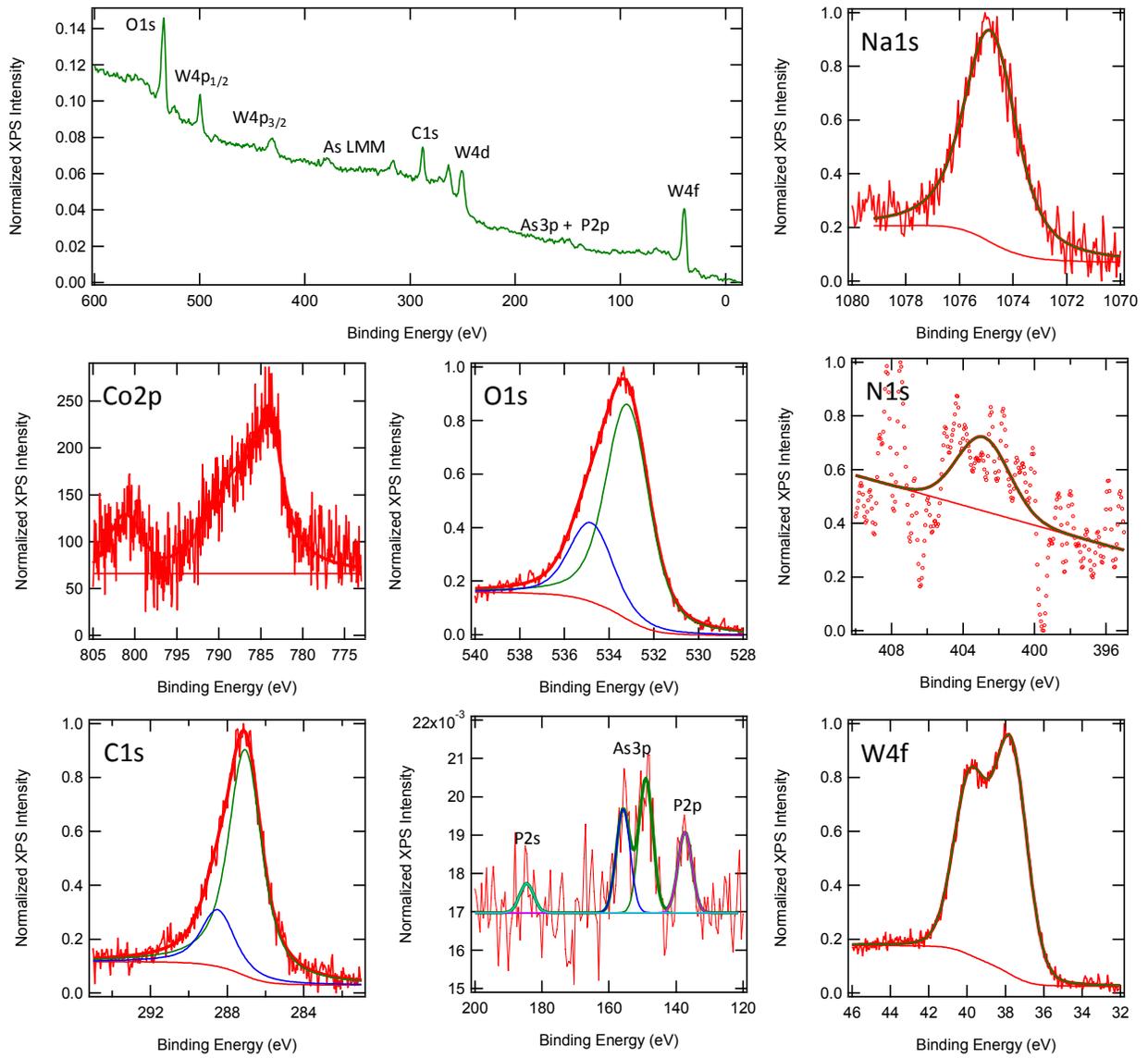

**Figure S18**. XPS spectra of **Na-2** powder.

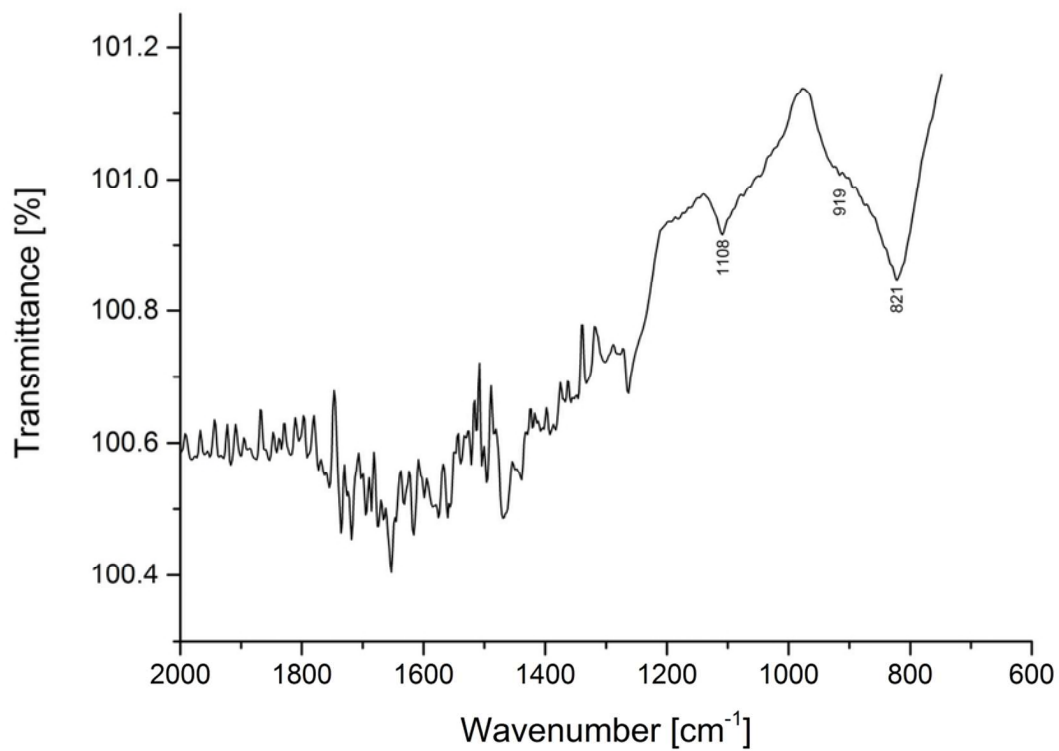

**Figure S19**. FT-IRRAS spectrum of layer obtained after overnight exposure of an Au plate to an aqueous solution of **Na-2**.

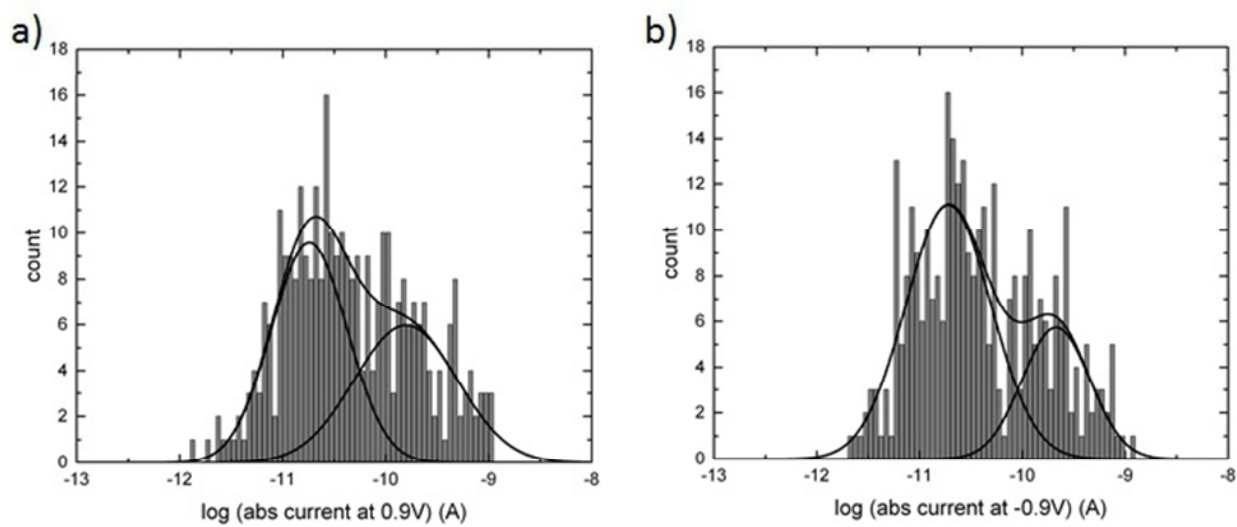

**Figure S20**. Current histograms at a given bias: +0.9V (**a**) and −0.9 V (**b**), as determined from *I-V* data in Fig. 4 (main text). The histograms are fitted by two log normal distributions with the following parameters: mean current/standard deviation −10.74 (i.e. $1.82 \cdot 10^{-11}$ A) /0.37 and −9.8 ($1.6 \cdot 10^{-10}$ A)/0.49 at 0.9 V, −10.72 ($1.9 \cdot 10^{-11}$ A)/0.41 and −9.67 ($2.1 \cdot 10^{-10}$ A) /0.33 at −0.9V.

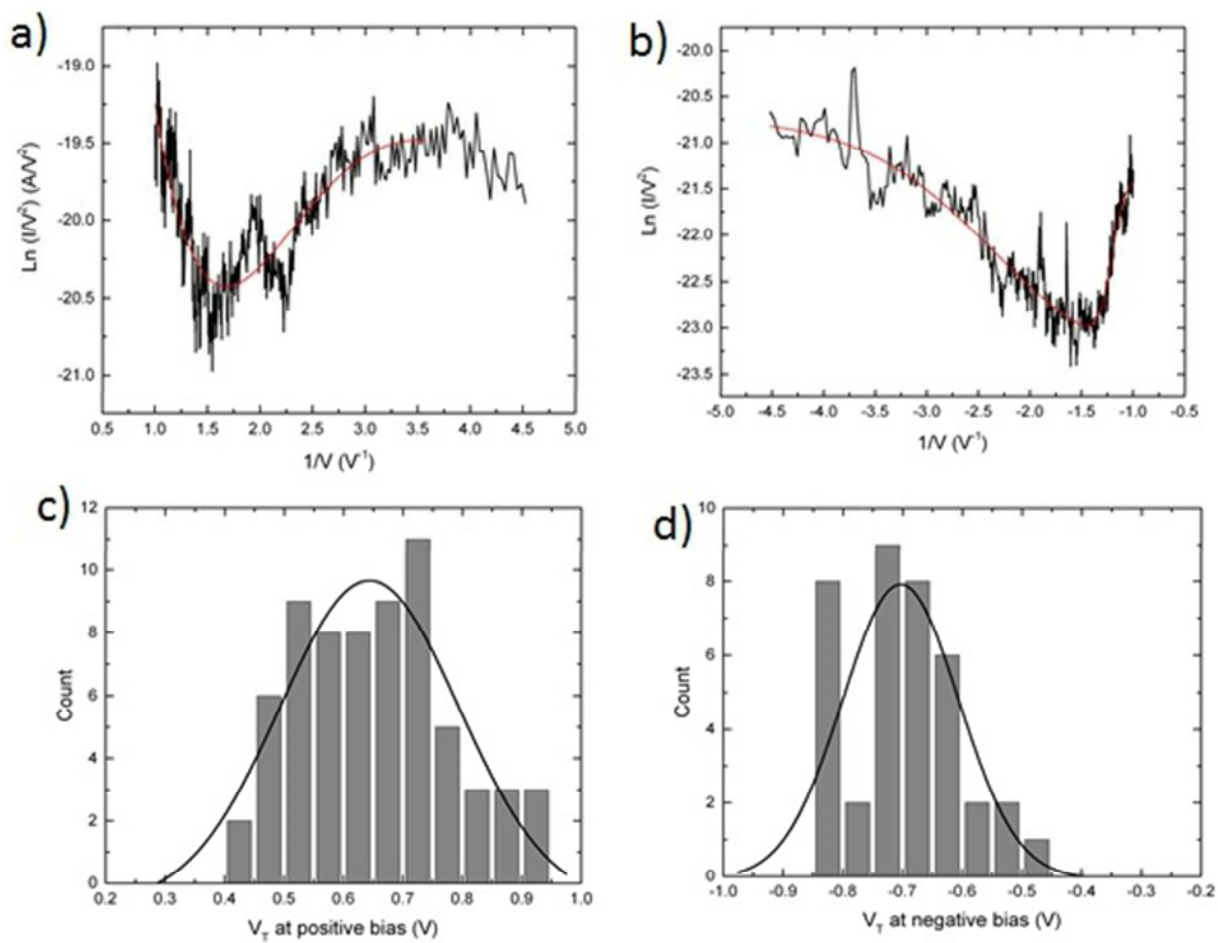

**Figure S21**. Typical TVS plots of the *I-V* curves at positive (**a**) and negative (**b**) voltages (red lines are guides to the eye) and histograms of the transition voltage $V_{T+}$ (67 counts) and $V_{T-}$ (38 counts) at positive (**c**) and negative (**d**) voltages, respectively (*I-V* data from Fig. 4, the counts for TVS is lower than the number of *I-V* curves because some *I-V* curves did not clearly show any minima in the TVS plot, too noisy curves). The $V_T$ absolute values are dispersed between ca. 0.4 and 0.9 V, centered at ca. 0.65 V given a position of the LUMO (see Eq. in Experimental part, main text) at 0.56 eV ± 0.17 eV. $V_T$ histograms are fitted with a Gaussian peak with the parameters: mean voltage/standard deviation of 0.64 V/0.15 V and −0.7 V/0.10 V, at positive and negative voltages, respectively.

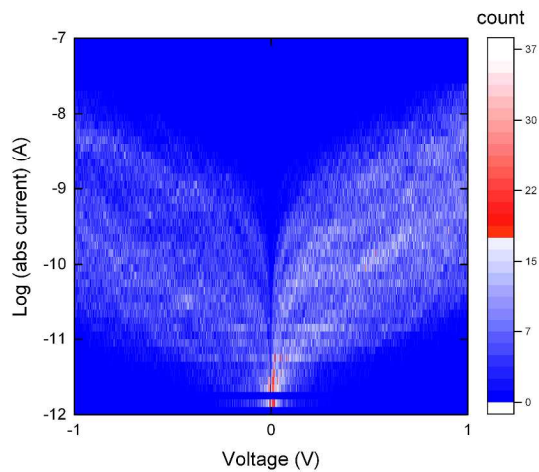

**Figure S22**. 2D current histogram of 872 *I-V* curves measured by C-AFM (at a loading force of 30 nN) on the SAM of **2** (second batch) chemically grafted on ultra-flat template stripped gold electrode (Au$^{TS}$). Voltages are applied on the Au$^{TS}$ electrode (C-AFM grounded).

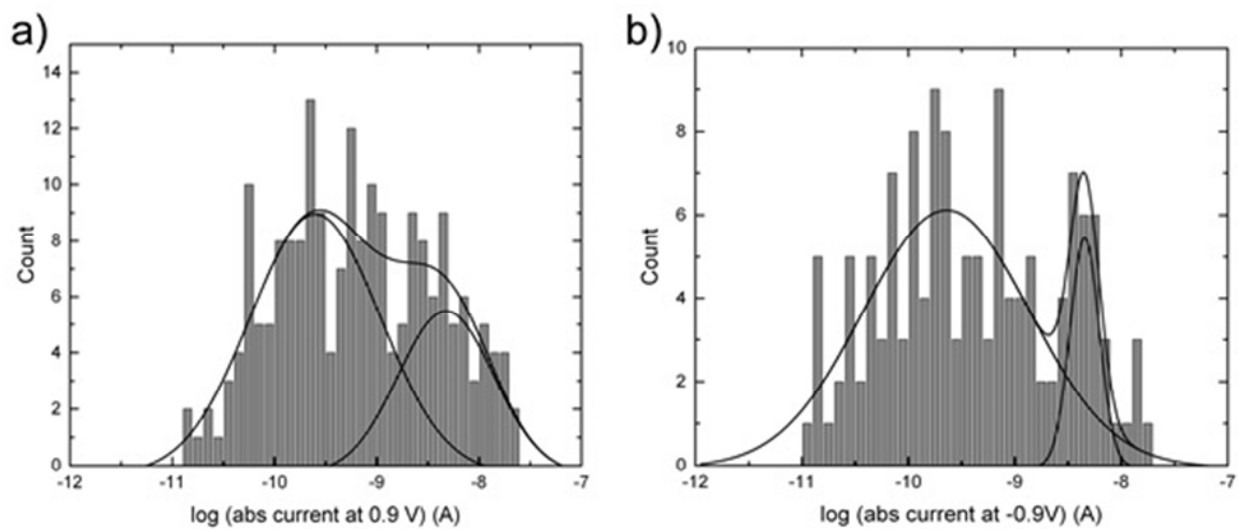

**Figure S23**. Current histograms at a given bias: +0.9V (a) and −0.9 V (b), from *I-V* data in Fig. S22. The histograms are fitted by two log normal distributions with the following parameters: mean current/standard deviation −9.6 (*i.e.* 2.5·10$^{-10}$ A) /0.6 and −8.32 (4.8·10$^{-9}$ A)/0.46 at 0.9 V, · 9.64 (2.3·10$^{-10}$ A)/0.77 and −8.34 (4.6·10$^{-9}$ A) /0.14 at −0.9V.

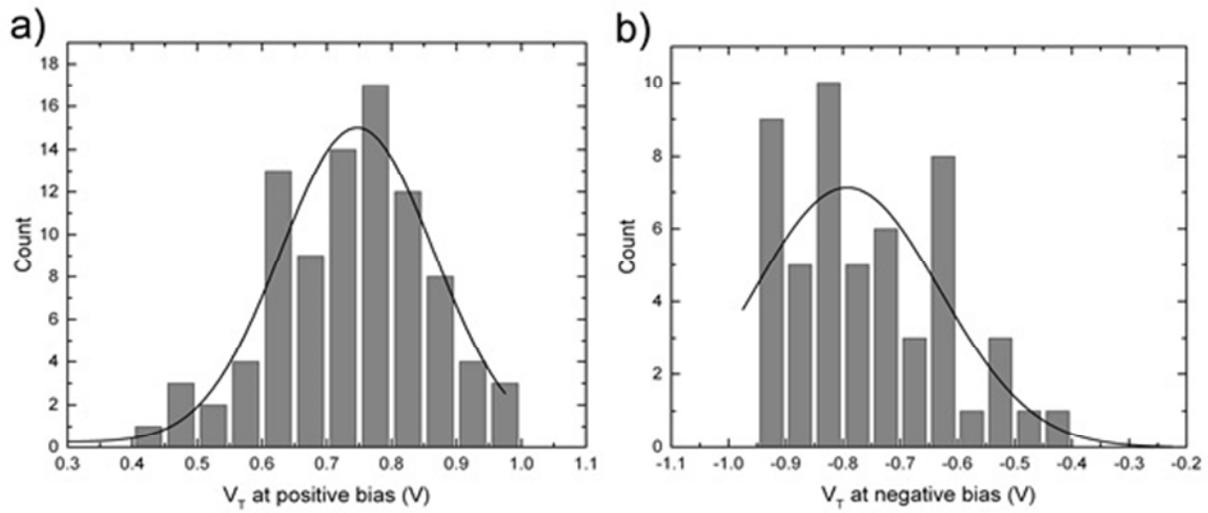

**Figure S24**. Histograms of the transition voltage $V_{T+}$ (90 counts) and $V_{T-}$ (52 counts) at positive (a) and negative (b) voltages from data Fig. S22, respectively. $V_T$ histograms are fitted with a Gaussian peak with the parameters: mean voltage/standard deviation of 0.75 V/0.11 V and −0.79 V/0.16 V, at positive and negative voltages, respectively.

**VII. Crystal packing of polyanions 1 in Na-1**

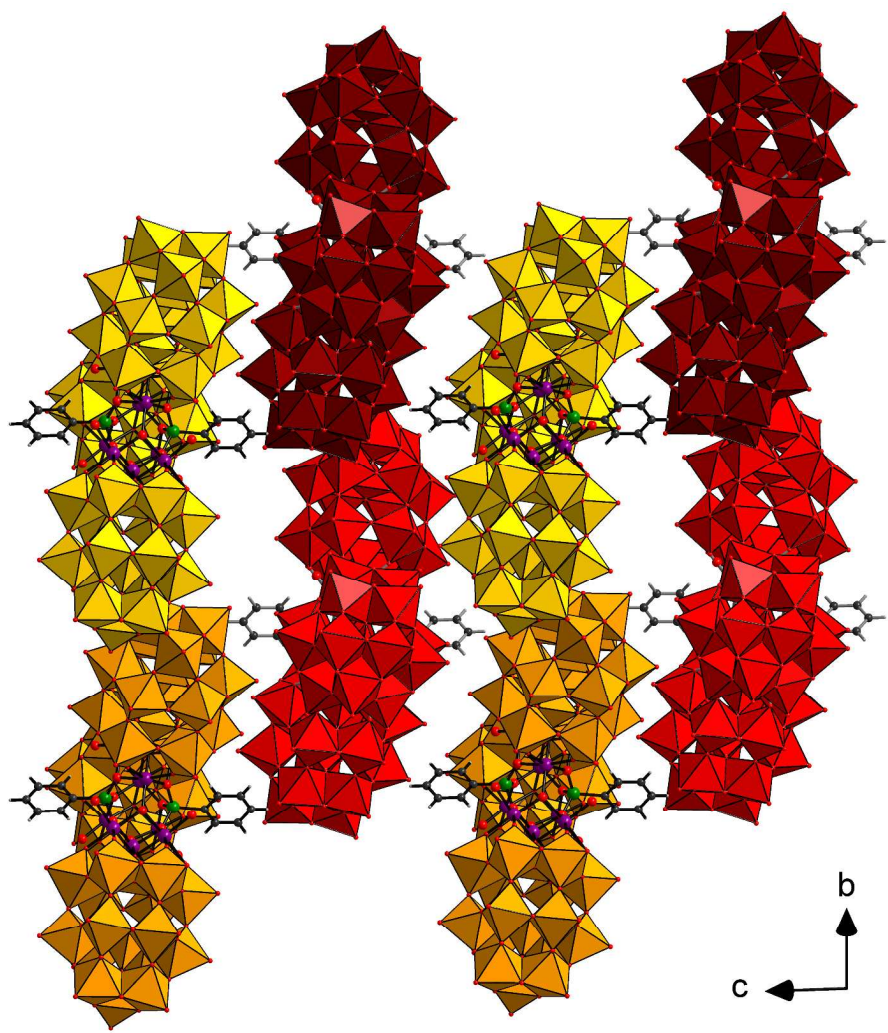

**Figure S25**. Packing of polyanions **1** in the crystal lattice of **Na-1**. View along the crystallographic *a* axis. The WO$_6$ and PO$_4$ polyhedra of the neighboring polyanions **1** alternate in color. Na and O atoms of crystal water are omitted for clarity. Color code for the other atoms: Co purple, O red, As green, C black, H gray spheres.

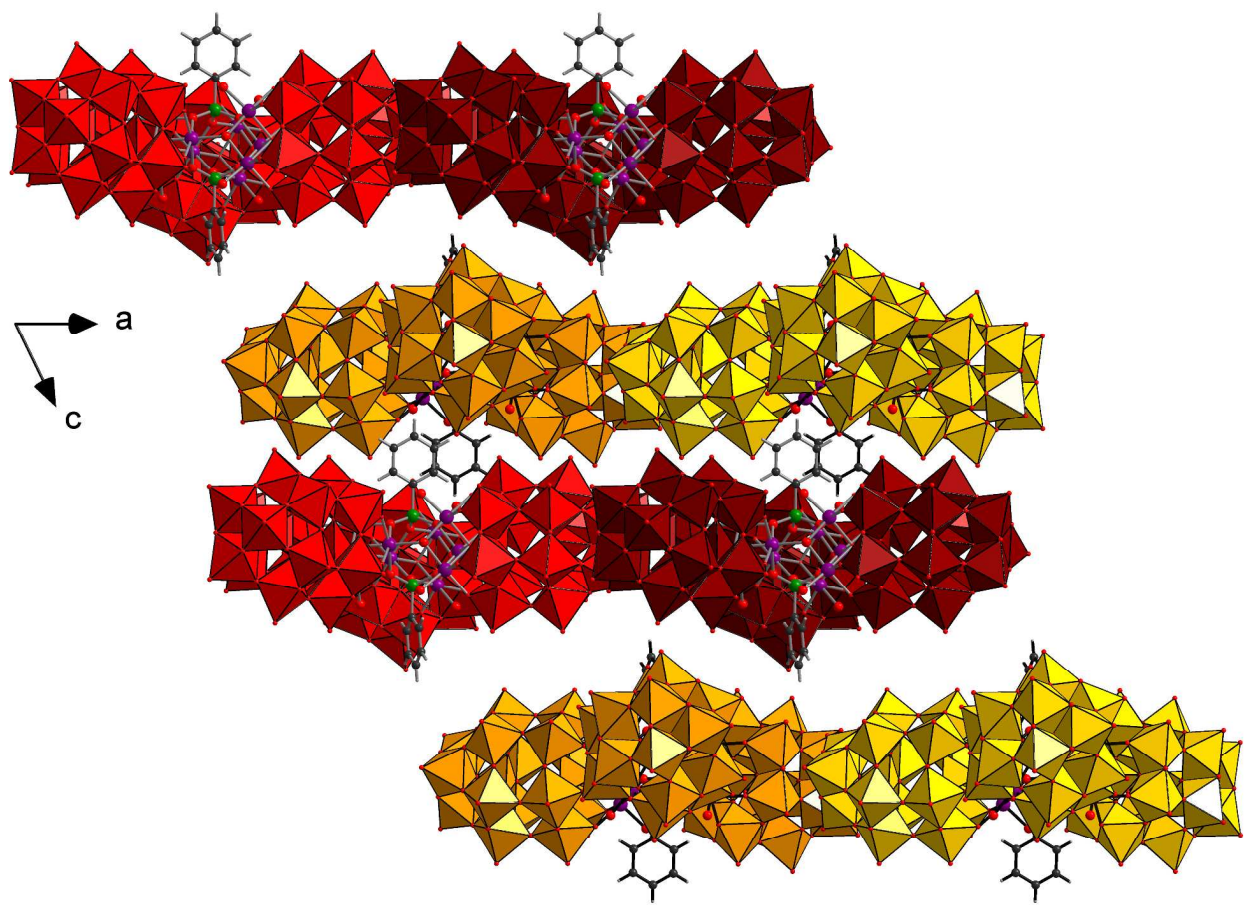

**Figure S26**. Packing of polyanions **1** in the crystal lattice of **Na-1**. View along *b*. Color code as in Fig. S25.

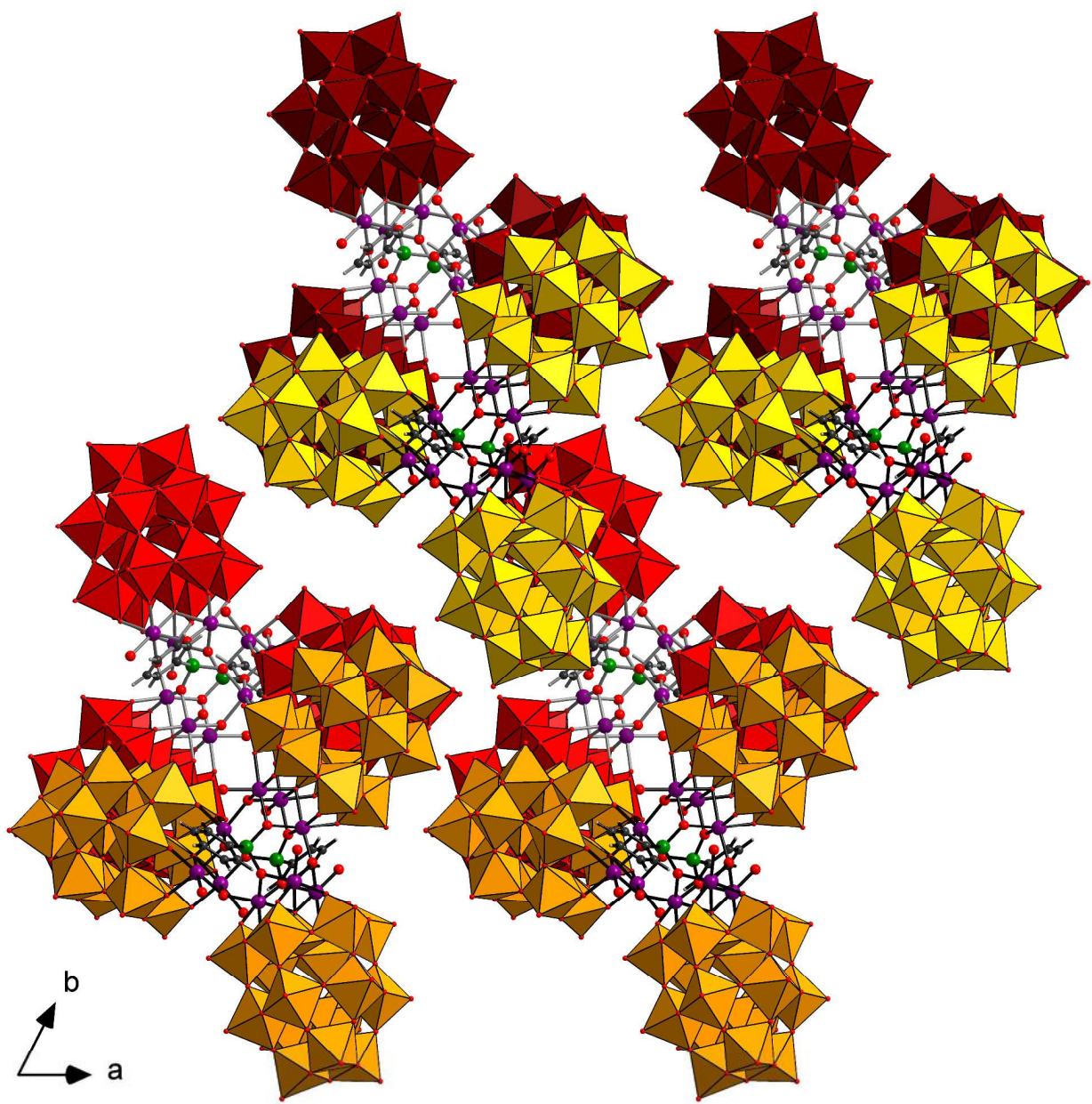

**Figure S27**. Packing of polyanions **1** in the crystal lattice of **Na-1**. View along *c*. Color code as in Fig. S25.

**VIII. Crystal packing of polyanions 2 in compound Na-2**

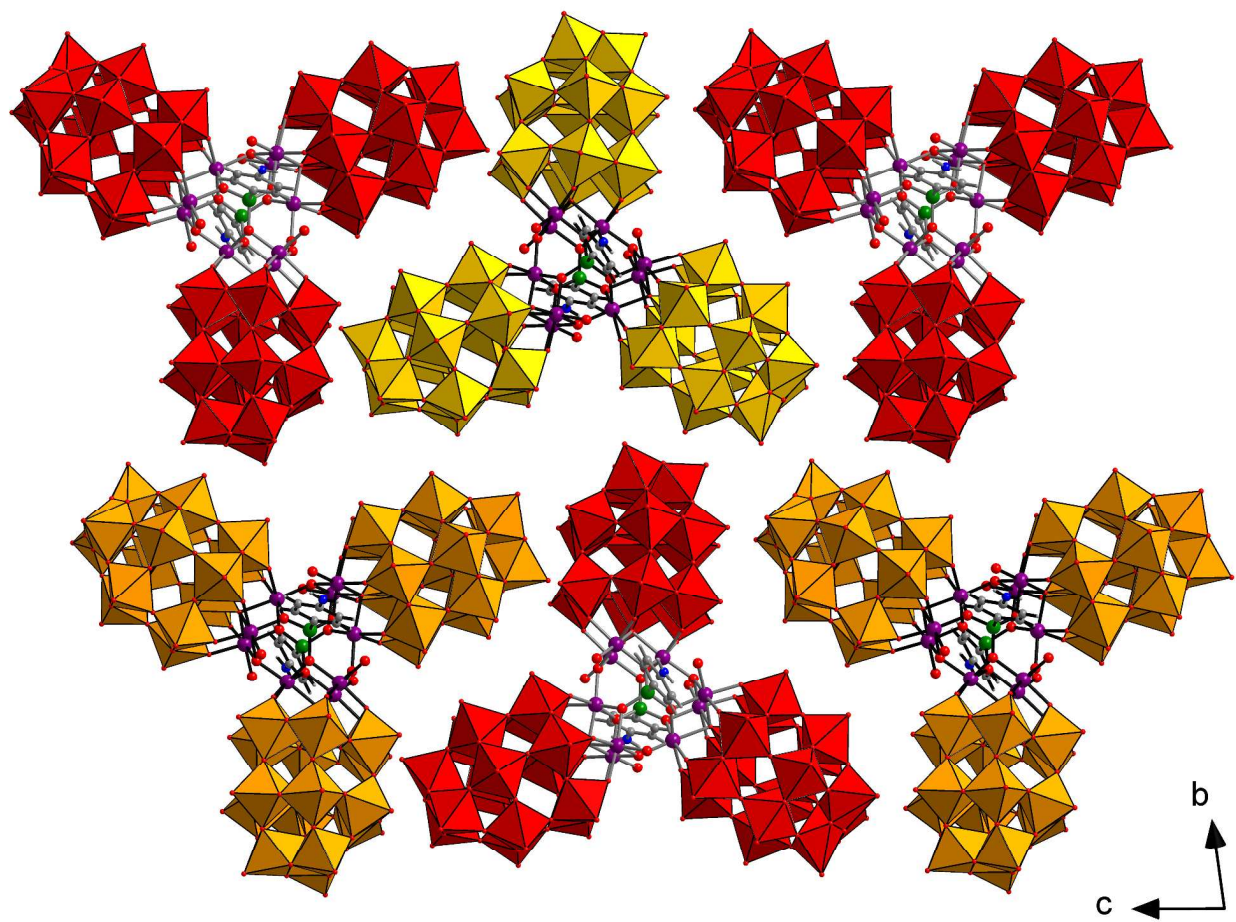

**Figure S28**. Packing of polyanions **2** in the crystal lattice of **Na-2**. View along *a*. The WO$_6$ and PO$_4$ polyhedra of the neighboring polyanions **2** alternate in color. Na and O atoms of crystal water are omitted for clarity. Color code for the other atoms: Co purple, O red, As green, C black, N blue, H gray spheres.

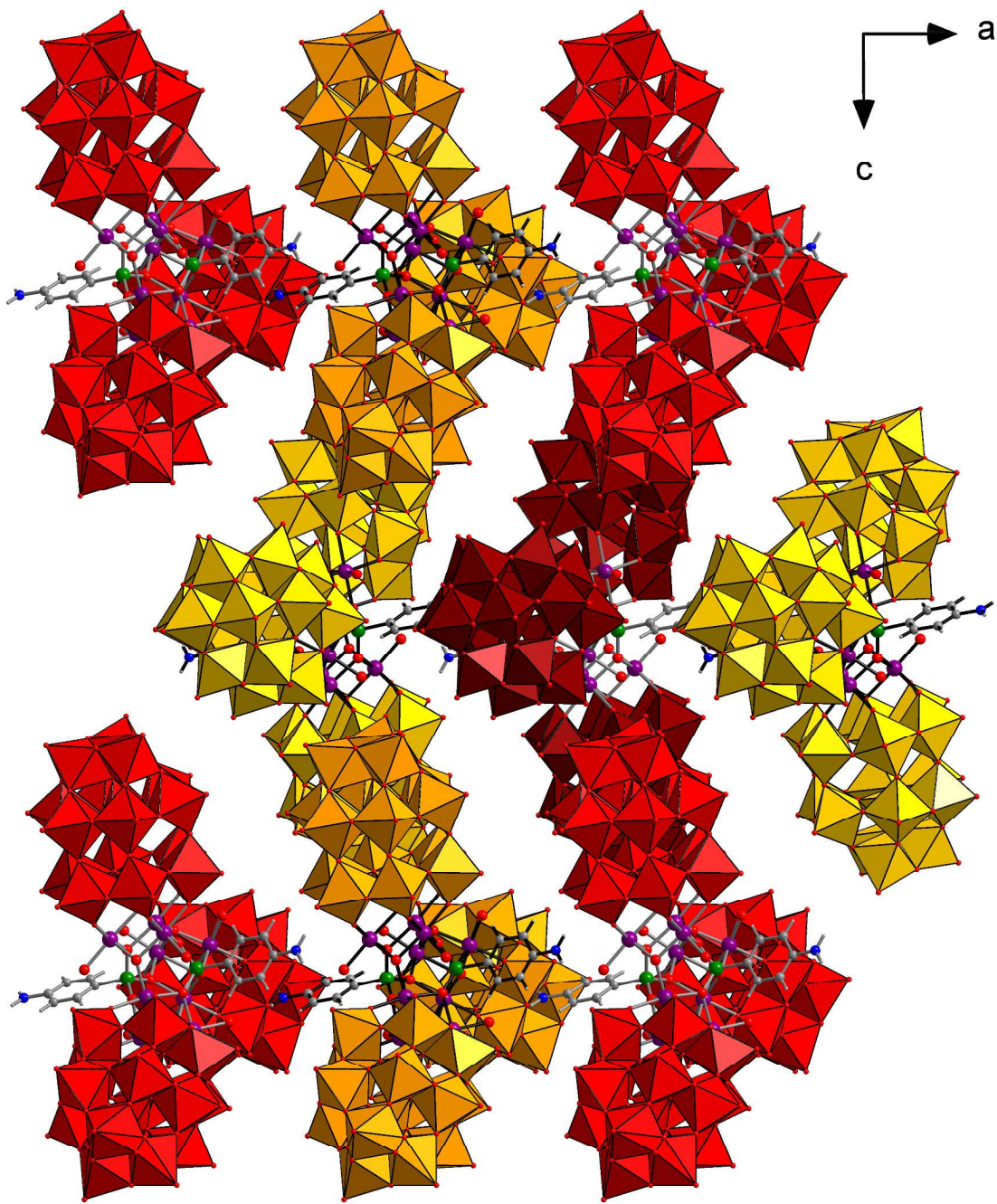

**Figure S29**. Packing of polyanions **2** in the crystal lattice of **Na-2**. View along *b*. Color code as in Fig. S28.

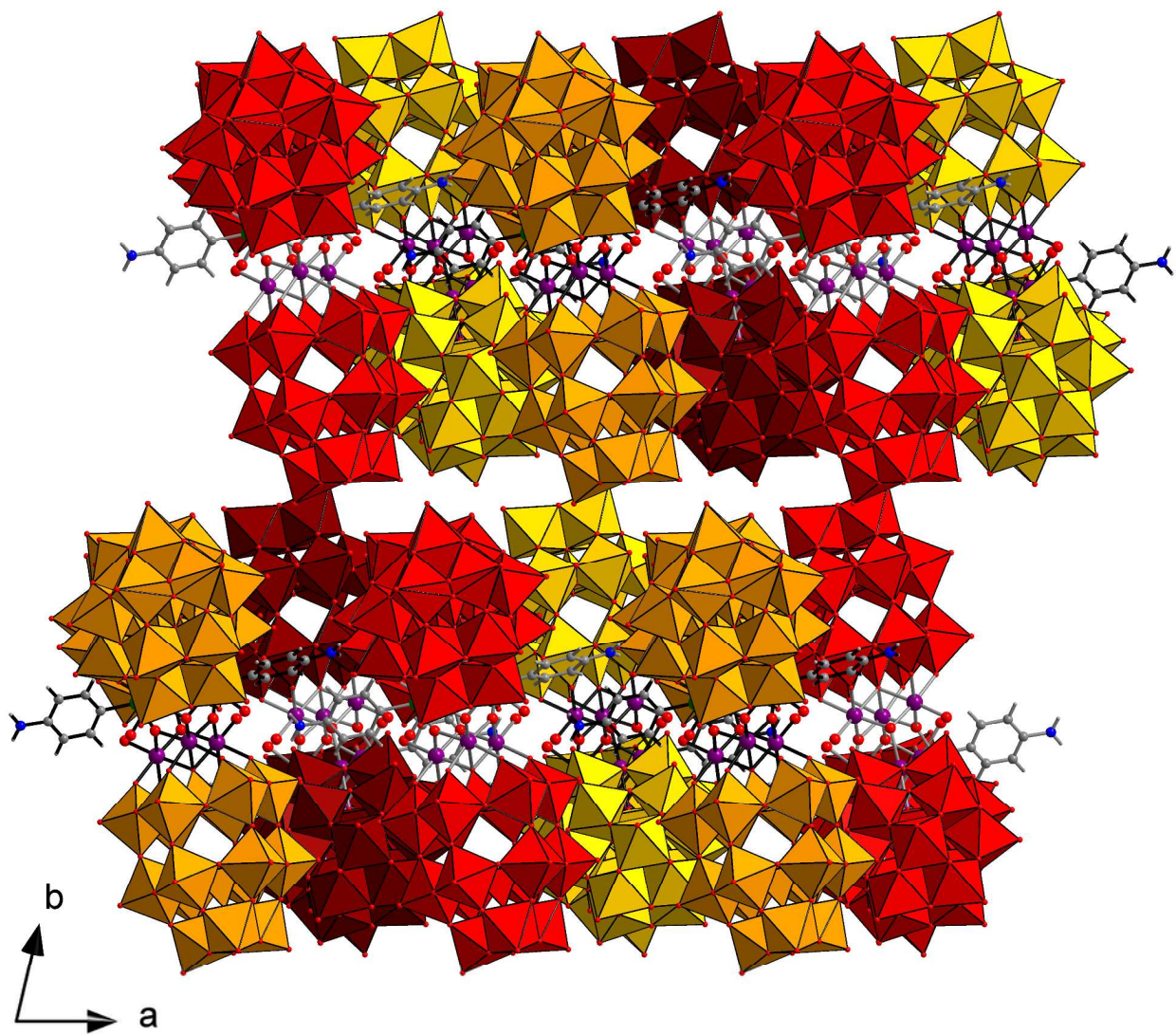

**Figure S30**. Packing of polyanions **2** in the crystal lattice of **Na-2**. View along *c*. Color code as in Fig. S28.